\documentclass[useAMS,usenatbib]{mn2e}
\usepackage{graphics}
\usepackage{epsf}
\usepackage{subfigure}
\def\ni{\noindent}

\begin{document}

\title[Cepheid PC \& AC relations III]{Period-color and amplitude-color relations in classical Cepheid variables III: The Large Magellanic Cloud Cepheid models }
\author[Kanbur \& Ngeow]{Shashi M. Kanbur$^{1}$\thanks{E-mail: kanbur@oswego.edu} and Chow-Choong Ngeow$^{2,3}$ 
\\
$^{1}$Department of Physics, State University of New York at Oswego, Oswego, NY 13126, USA
\\
$^{2}$Department of Astronomy, University of Massachusetts, Amherst, MA 01003, USA
\\
$^{3}$Department of Astronomy, University of Illinois, Urbana-Champaign, IL 61801, USA
}

\date{Accepted 2005 month day. Received 2005 month day; in original form 2005 June 05}

%\pagerange{\pageref{1}--\pageref{15}} \pubyear{2003}

\maketitle

\begin{abstract}

Period-colour (PC) and amplitude-colour (AC) relations are studied for the Large Magellanic Cloud (LMC) Cepheids under the theoretical framework of the hydrogen ionization front (HIF) - photosphere interaction. LMC models are constructed with pulsation codes that include turbulent convection, and the properties of these models are studied at maximum, mean and minimum light. As with Galactic models, at maximum light the photosphere is located next to the HIF for the LMC models. However very different behavior is found at minimum light. The long period ($P>10$days) LMC models imply that the photosphere is disengaged from the HIF at minimum light, similar to the Galactic models, but there are some indications that the photosphere is located near the HIF for the short period ($P<10$ days) LMC models. We also use the updated LMC data to derive empirical PC and AC relations at these phases. Our numerical models are broadly consistent with our theory and the observed data, though we discuss some caveats in the paper. We apply the idea of the HIF-photosphere interaction to explain recent suggestions that the LMC period-luminosity (PL) and PC relations are non-linear with a break at a period close to 10 days. Our empirical LMC PC and PL relations are also found to be non-linear with the $F$-test. Our explanation relies on the properties of the Saha ionization equation, the HIF-photosphere interaction and the way this interaction changes with the phase of pulsation and metallicity to produce the observed changes in the LMC PC and PL relations. 

\end{abstract}


%************************************
%  INTRODUCTION
%************************************

\section{Introduction}

     \citet{cod47} found that the Galactic Cepheids follow a spectral type that is independent of their pulsational periods at maximum light and gets later as the periods increase at minimum light. \citet[][hereafter SKM]{sim93} used radiative hydrodynamical models to explain these observational phenomena as being due to the location of the hydrogen ionization front (HIF) relative to the photosphere. Their results agreed very well with Code's observation. SKM further used the Stefan-Boltzmann law applied at the maximum and minimum light, together with the fact that radial variation is small in the optical \citep{cox80}, to derive:

     \begin{eqnarray}
       \log T_{max} - \log T_{min} = {1\over{10}}(V_{min} - V_{max}),
     \end{eqnarray}

     \ni where $T_{max/min}$ are the effective temperature at the maximum/minimum light, respectively. If $T_{max}$ is independent of the pulsation period $P$ (in days), then equation (1) predicts there is a relation between the $V$-band amplitude and the temperature (or the colour) at minimum light, and vice versa. In other words, if the period-colour (PC) relation at maximum (or minimum) light is flat, then there is an amplitude-colour (AC) relation at minimum (or maximum) light. Equation (1) has shown to be valid theoretically and observationally for the classical Cepheids and RR Lyrae variables \citep{kan04,kan05}.  

     For the RR Lyrae variables, \citet{kan95} and \citet{kan96} used linear and non-linear hydrodynamic models of RRab stars in the Galaxy to explain why RRab stars follow a flat PC relation at {\it minimum} light. Later, \citet{kan05} used MACHO RRab stars in the LMC to prove that LMC RRab stars follow a relation such that higher amplitude stars are driven to cooler temperatures at maximum light. Similar studies were also carried out for Cepheid variables, as in SKM, \citet{kan96}, \citet[][hereafter Paper I]{kan04} and \citet[][hereafter Paper II]{kan04a}. In contrast to the RR Lyrae variables, Cepheids show a flat PC relation at the {\it maximum} light, and there is a AC relation at the minimum light. Therefore, the PC relation and the AC relation are intimately connected. All these studies are in accord with the predictions of equation (1).
     
     In Paper I, the Galactic, Large Magellanic Cloud (LMC) and Small Magellanic Cloud (SMC) Cepheids were analyzed in terms of the PC and AC relations at the phase of maximum, mean and minimum light. One of the motivations for this paper originates from recent studies on the non-linear LMC PC relation \citep[as well as the period-luminosity, PL, relation. See Paper I;][]{tam02a,san04,nge05}: the optical data are more consistent with two lines of differing slopes which are continuous or almost continuous at a period close to 10 days. Paper I also applied the the $F$-test \citep{wei80} to the PC and AC relations at maximum, mean and minimum $V$-band light for the Galactic, LMC and SMC Cepheids. The $F$-test results implied that the LMC PC relations are broken or non-linear, in the sense described above, across a period of 10 days, at mean and minimum light, but only marginally so at maximum light. The results for the Galactic and SMC Cepheids are similar, in a sense that at mean and minimum light the PC relations do not show any non-linearity and the PC(max) relation exhibited marginal evidence of non-linearity. For the AC relation, Cepheids in all three galaxies supported  the existence of two AC relations at maximum, mean and minimum light. In addition, the Cepheids in these three galaxies also exhibited evidence of the PC-AC connection, as implied by equation (1), which give further evidence of the HIF-photosphere interactions as outlined in SKM.

     To further investigate the connection between equation (1) and the HIF-photosphere interaction, and also to explain Code's observations with modern stellar pulsation codes, Galactic Cepheid models were constructed in Paper II. In contrast to SKM's purely radiative models, the stellar pulsation codes used in Paper II included the treatment of turbulent convection as outlined in \citet{yec98}. One of the results from Paper II was that the general forms of the theoretical PC and AC relation matched the observed relations well. The properties of the PC and AC relations for the Galactic Cepheids with $\log(P)>0.8$ can be explained with the HIF-photosphere interaction. This interaction, to a large extent, is independent of the pulsation codes used, the adopted ML relations, and the detailed input physics. 
   
     The aim of this paper is to extend the investigation of the connections between PC-AC relations and the HIF-photosphere interactions in theoretical pulsation models of LMC Cepheids, in addition to the Galactic models presented in Paper II. In Section 2, we describe the basic physics of the HIF-photosphere interaction. The updated observational data, after applying various selection criteria, that used in this paper are described in Section 3. In Section 4, the new empirical PC and AC relations based on the data used are presented. In Section 5, we outline our methods and model calculations, and the results are presented in Section 6. Examples of the HIF-photosphere interaction in astrophysical applications are given in Section 7. Our conclusions \&  discussion are presented in Section 8. Throughout the paper, short and long period Cepheid are referred to Cepheids with period less and greater than 10 days, respectively.

\section{The Physics of HIF-Photosphere Interactions}

     The partial hydrogen ionization zone (or the HIF) moves in and out in the mass distribution as the star pulsates. It is possible that the HIF will interact with the photosphere, defined at optical depth ($\tau$) of 2/3, at certain phases of pulsation. For example, SKM suggested that this happened at maximum light for the Galactic Cepheids, as the HIF is so far out in the mass distribution that the photosphere occurs right at the base of the HIF. The sharp rise of the opacity wall (where the mean free path goes to zero) due to the existence of HIF prevents the photosphere moving further into the mass distribution and hence erases any ``memory'' of global stellar conditions, including the underlying PC relation. This lead to a flat relation between period \& temperature, period \& colour and period \& spectral type at maximum light, as seen in SKM and Paper II. At other phases, since the HIF does not interact with the photosphere, the temperature of the star (or the colour) follows the underlying global PC relation.

     The HIF-photosphere interaction also relies on the properties of the Saha ionization equation and the structural properties of the outer envelopes of Cepheids. It is well known that the partition functions in the Saha ionization equation are formally divergent unless some atomic physics is used to truncate them. In the pulsation codes we used, we approximate the partition functions of various atoms by their ground state statistical weights. The properties of the Saha ionization equation in Cepheid envelopes are such that hydrogen starts to ionize at a temperature that is almost independent of density, for a certain range of low densities. Outside of this range of density, the density dependence increases. Thus, when the photosphere is very close to, or engaged with the HIF and the density of these regions is reasonably low, the temperature of the photosphere is less dependent on the surrounding density and hence the global stellar parameters. At higher densities, the temperature at which hydrogen ionizes becomes more sensitive to density and hence more sensitive to global stellar parameters.

     If the photosphere is far from the HIF, or disengaged, then the location of the photosphere and hence the temperature of the photosphere, is again strongly dependent on density and hence on global stellar parameters. That is why the photosphere needs to be close to, or engaged with the HIF for this effect to take place. Moreover, this dependence on density is not sharp so that for  "low" and "high" densities the density dependence of the photospheric temperature is weak and strong respectively. An examination of figure 15.1 in \citet{cox68} demonstrates that this is plausible. Thus as the star pulsates, the photospheric temperature has a density dependence that can be strong or weak depending on phase. An example where the density dependence is weak are the Galactic long period Cepheids at maximum light (SKM, Paper II): these Cepheids display a flat PC relation at maximum light. These properties of the HIF-photosphere interaction can, in turn, affect the temperature of the photosphere and hence the colour of the Cepheid.
     
    Here we investigate the idea that LMC Cepheids with periods below 10 days are such that the HIF and photosphere are engaged through most of the pulsation cycle. At periods greater than 10 days, the photosphere only engages with the HIF at maximum light. The transition is sharp because the photosphere is either at the base of the HIF or it is not. The transition occurs because as the period increases, the $L/M$ ratio increases and this implies the HIF is located further inside in the mass distribution, changing the phase at which it can interact with the photosphere \citep{kan95}. The structure of Galactic Cepheids is such that this interaction only occurs at maximum light, even for Cepheids with periods shorter than 10 days.

\section{Updated LMC Cepheid Data}

     In Paper I, we constructed the light curves of fundamental mode Cepheids in the LMC by using the extensive photometric dataset in the OGLE (Optical Gravitational Lensing Experiment) database. However, the dataset used in Paper I was downloaded in 2002, prior to the updated version of the dataset that was available after April 24, 2003 (OGLE website, Udalski 2004 [private communication]). The updated version includes additional $V$- and $I$-band data for most of the Cepheids. In addition, the periods have been refined  by the OGLE team using the complete set of photometric data. Due to these reasons, we decided to repeat the light curve construction \citep{nge03} with the updated data and periods. Since the Cepheids in the OGLE database are truncated at $\log(P)\sim1.5$, due to the saturation of the CCD detector for the longer period (hence brighter) Cepheids \citep{uda99}, we include some additional LMC Cepheid data from \citet{mof98}, \citet{bar99} and \citet{seb02} to extend the period coverage to $\log(P)>1.5$ in our sample. The requirements that govern our choice of the published photometric data are: (a) latest observations that use the modern day CCD cameras; (b) high quality data with large number of data points per light curve, which provide uniform phase coverage and small scatter of the light curve; and (c) as homogeneous as possible (i.e., from a minimal number of sources) to avoid any additional systematic errors. These requirements are essential to construct accurate light curves to allow the estimation of colours and magnitudes at maximum, mean and minimum light for our PC and AC study. Hence we did not include some of the older photometric data in this study. 
     
     The photometric data of all Cepheids, comprising 771 from OGLE database, 14 from \citet{mof98}+\citet{bar99} and 39 from \citet{seb02}\footnote{For \citet{seb02} data, the photometry data are only available for the long period Cepheids. To avoid duplication, we remove the long period Cepheids in \citet{seb02} that are labeled ``Ogle'', those \citet{seb02} identified as the OGLE Cepheids, and 5 ``HV'' Cepheids that are the same in \citet{mof98}. This left 39 ``HV'' Cepheids from \citet{seb02} sample.}, were mainly fit with $n=4$ to $n=8$ Fourier expansions ($n$ is the order of Fourier expansion) using the simulated annealing method described in \citet{nge03} to the $V$- and $I$-band photometric data. This is in contrast to Paper I that only applied $n=4$ Fourier fits. However, for some of the OGLE long period Cepheids ($P>11.5$ days), it was found out that the quality of the fitted light curves could be improved by using a higher order Fourier expansion, hence we extended the fit to $n=12$ for these long period Cepheids. All the fitted light curves were visually inspected and the best-fit light curves from the different orders of the Fourier expansions were selected. To the best of our knowledge, this analysis also represents a major improvement in the Fourier analysis of the OGLE data. The extinction is corrected with the standard procedure, i.e. $(V-I)_0=(V-I)-(R_V-R_I)\times E(B-V)$ with $R_V=3.24$ and $R_I=1.96$ \citep{uda99b}. The values of $E(B-V)$ for each OGLE Cepheids are taken from the OGLE database \citep{uda99b}, while for the Cepheids in \citet{mof98}+\citet{bar99} and in \citet{seb02}, the values of $E(B-V)$ are adopted from \citet{san04} and/or \citet{per04}. 

     To guard against some ``bad'' Cepheids or other contamination in our sample, and select only the good Cepheids in both bands, we removed some Cepheids in the sample according to the following criteria \citep[see also][]{kan03}:

%************************************************
%  FIGURE: PC MAX/MIN outliers
%************************************************

     \begin{figure}
       \centering 
       \epsfxsize=7.5cm{\epsfbox{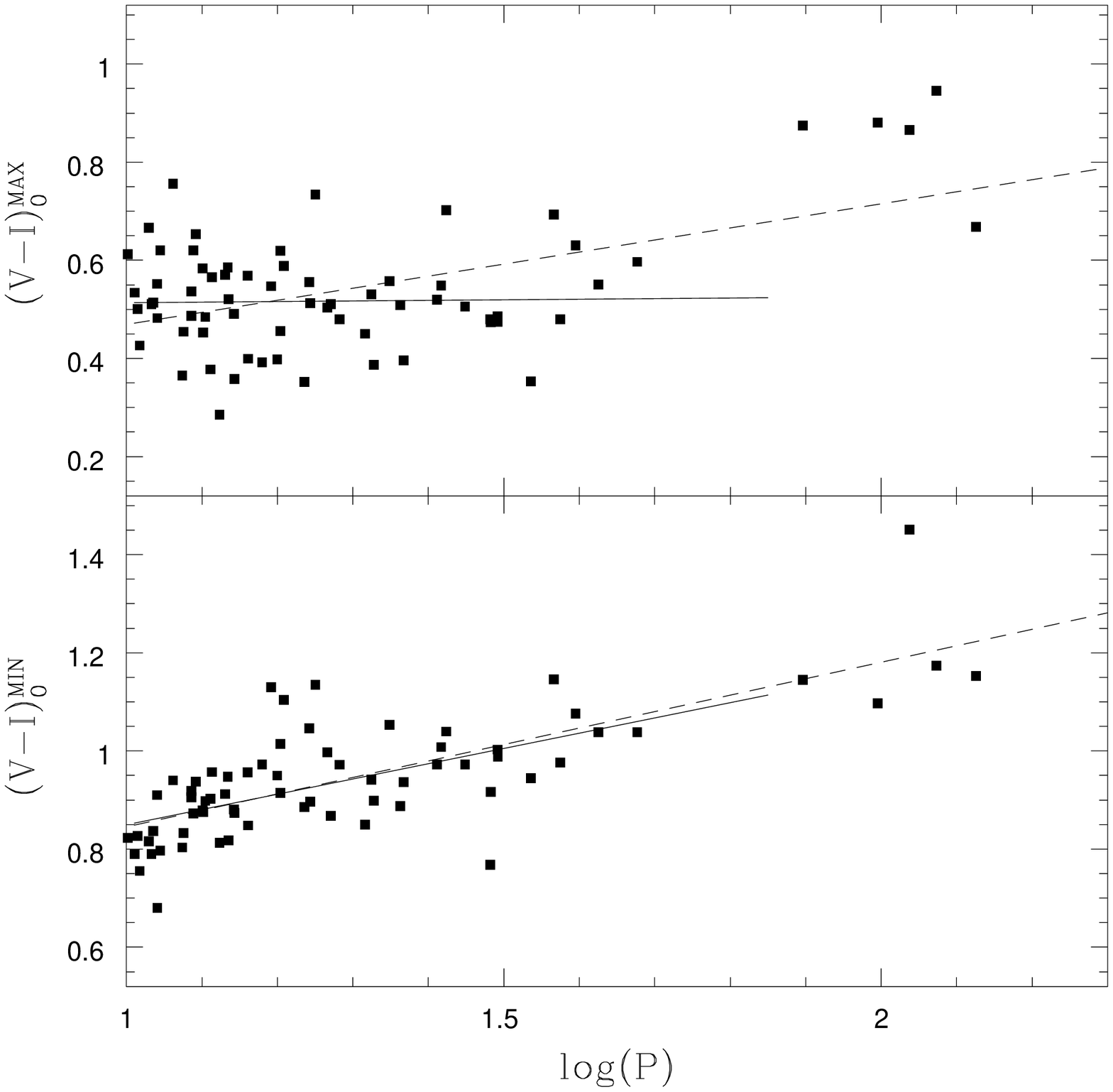}}
       \caption{The PC relation at maximum (upper panel) and minimum (lower panel) light for the long period Cepheids in our sample (after removal of outliers). The solid and dashed lined are the regressions that exclude and include the longest period Cepheids. It is clear that including the Cepheids with $\log(P)>1.8$ has made the PC(max) relation becomes steeper, even though the two regressions are consistent to each others at the minimum light.}
       \label{figpcmax}
     \end{figure}

\begin{figure*}
       \vspace{0cm}
       \hbox{\hspace{0.2cm}\epsfxsize=7.5cm \epsfbox{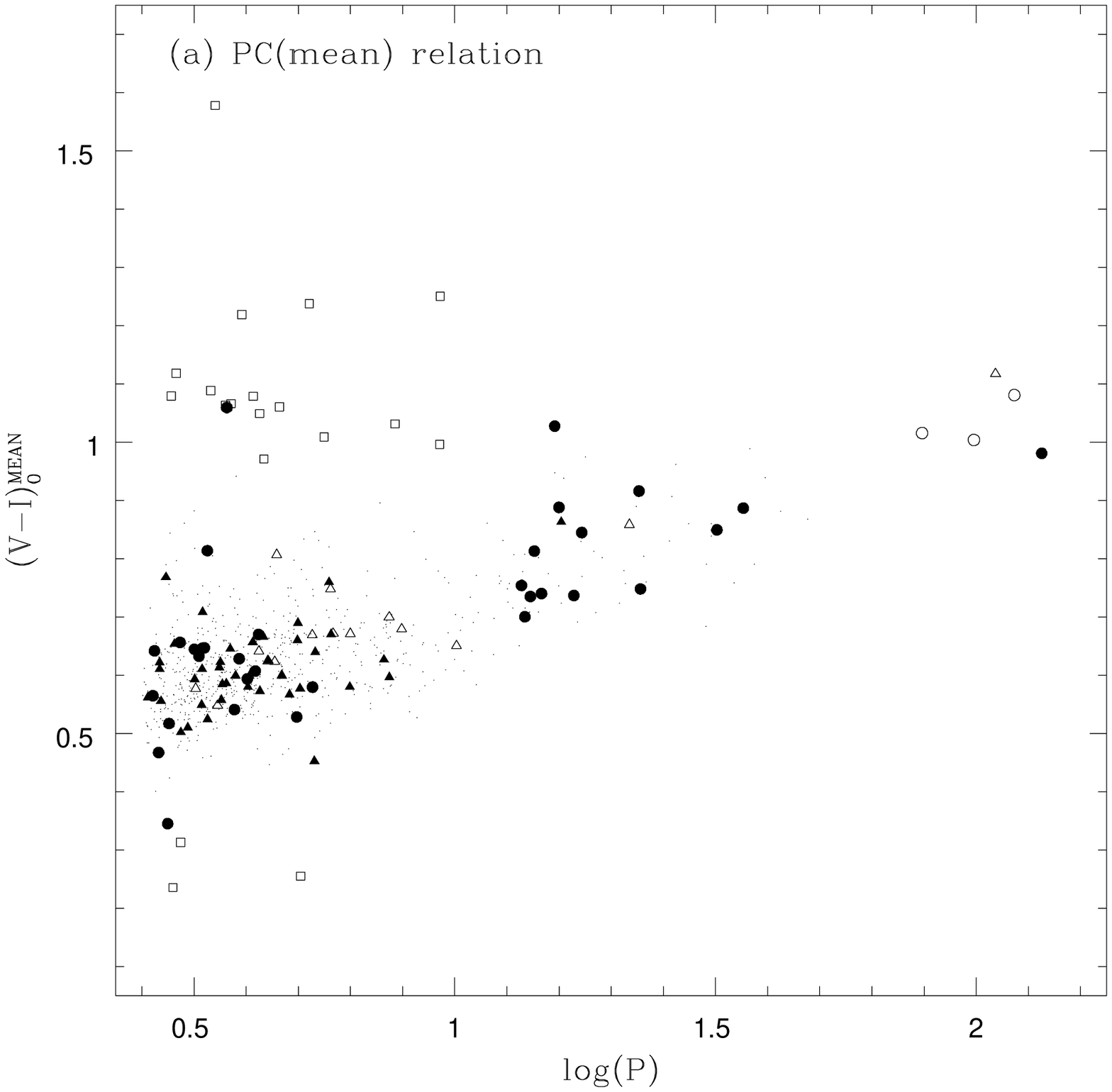}
         \epsfxsize=7.5cm \epsfbox{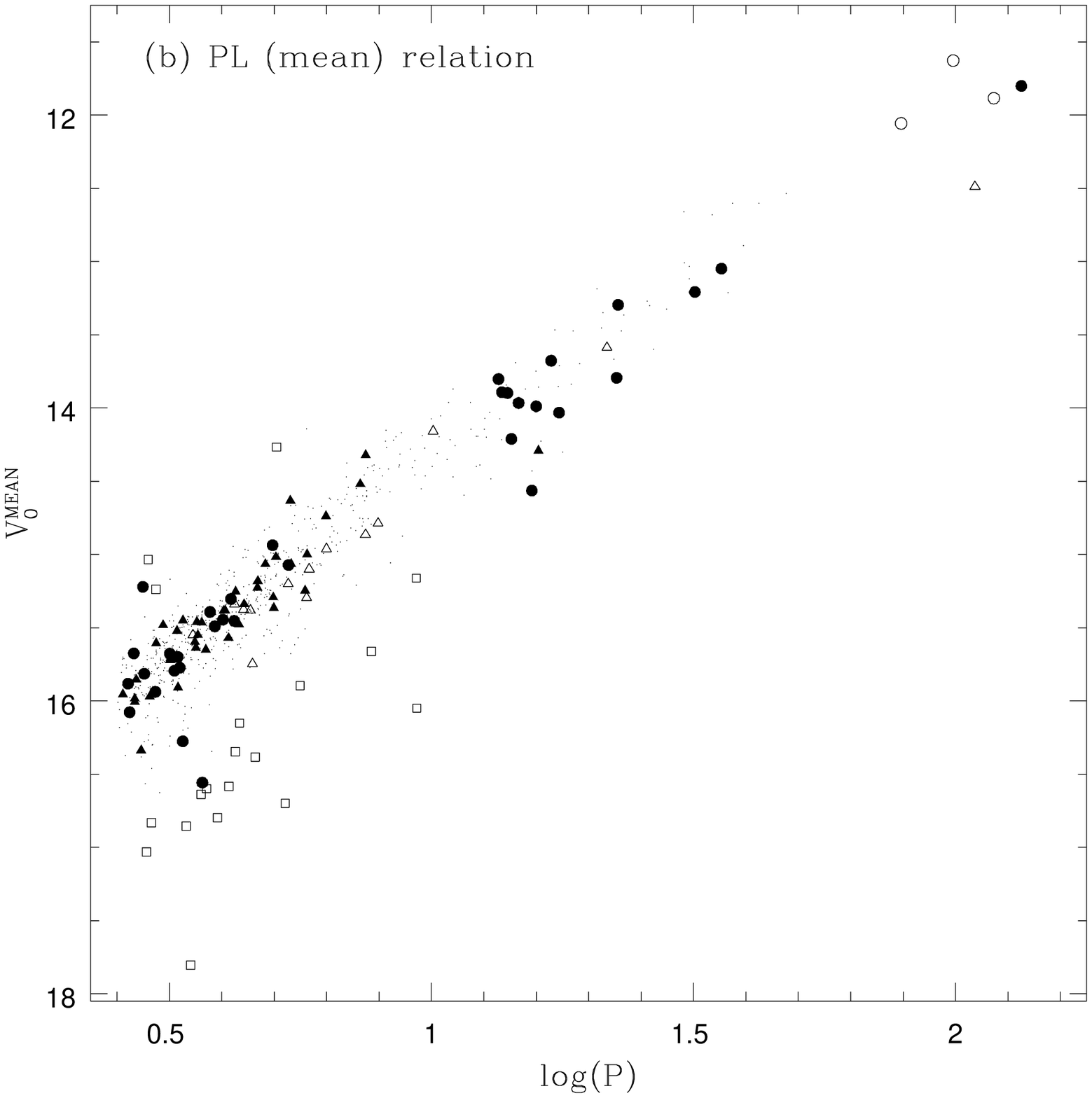}}
       \hbox{\hspace{0.2cm}\epsfxsize=7.5cm \epsfbox{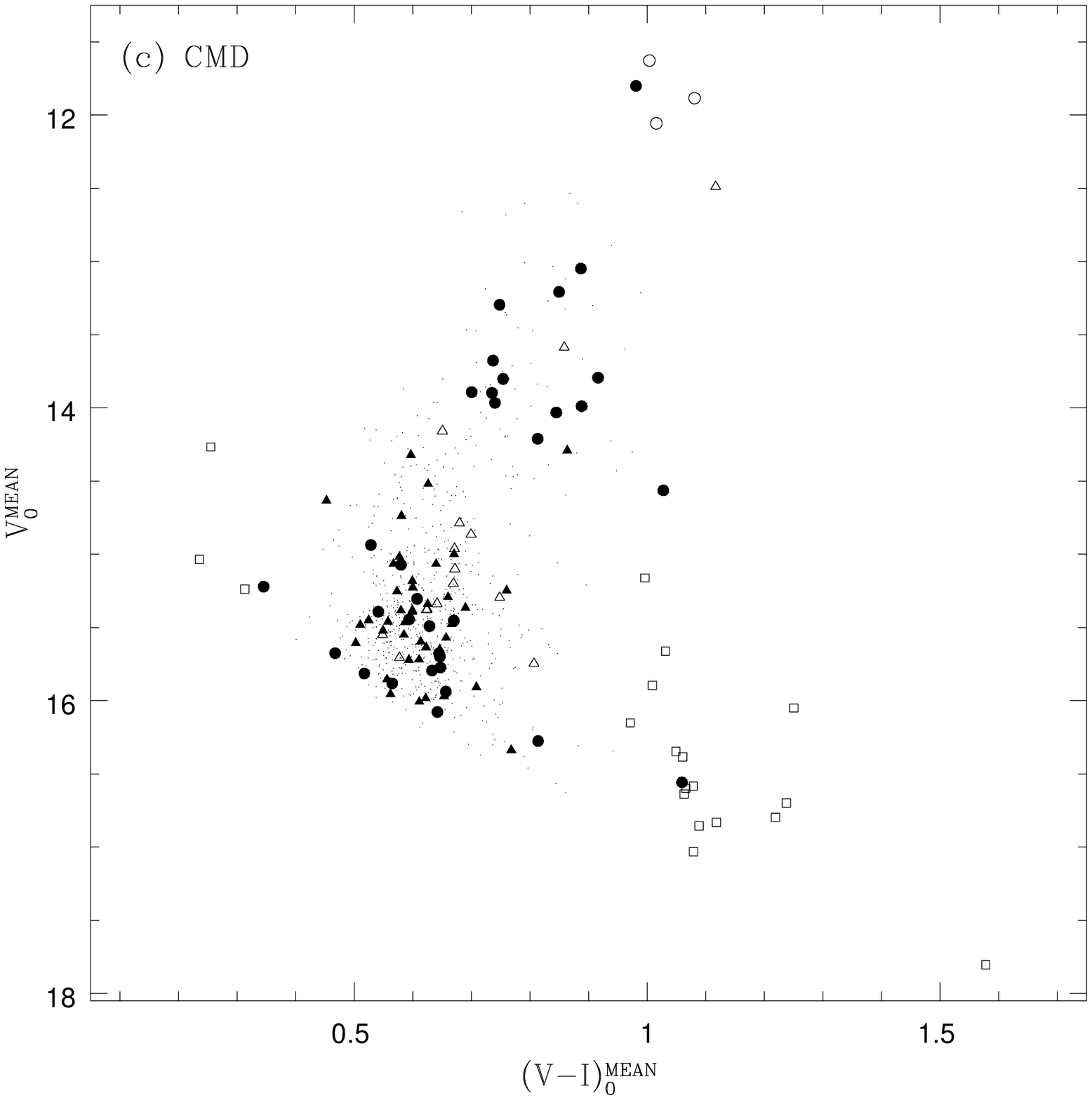}
         \epsfxsize=7.5cm \epsfbox{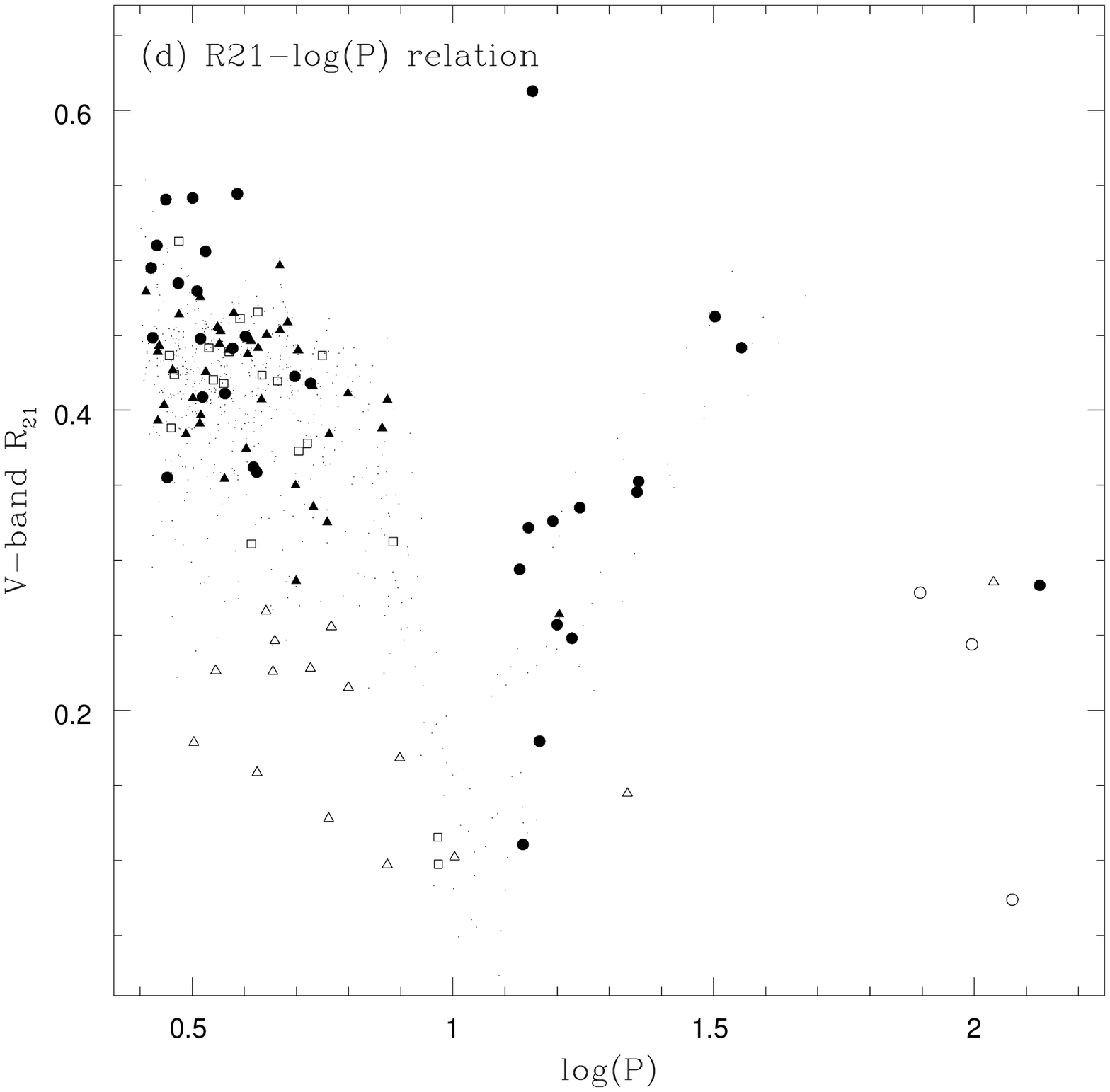}}
       \vspace{0cm}
       \caption[Locations of the LMC outliers in various relations and CMD]{--The various relations and the colour-magnitude diagram (CMD) for the LMC Cepheids, after removing Cepheids with $\log(P)<0.4$ and those lacking of $V$-band photometry. The symbols are (see text for details): filled circles = Cepheid with poor fit of the light curves; open squares = the rejected Cepheids after colour-cut of $0.35<(V-I)^{\mathrm{mean}}_0<0.95$; open triangles = the excluded Cepheids after the amplitude cut; filled triangles = the removed Cepheids for possible duplicity; and open circles = the longest period Cepheids, which are the outliers in PC(max) relation. The dots are the remaining good Cepheids in our sample.}
       \label{figcut}
     \end{figure*}

     \begin{enumerate}

     \item Cepheids without $V$- and/or $I$-band photometry, or the number of data per light curve (in either bands or both) is too low to fit a $n=4$ Fourier expansion.

     \item Cepheids with poorly fitted or unacceptable $V$- and/or $I$-band light curves in the sample, such as those with a large scatter of data points or with bad-phase coverage (large gaps between the phased data points). Most of the magnitudes, as well as the colours, at the maximum and/or minimum light from these fitted light curves are very uncertain.

     \item Cepheids with possible duplicity in the OGLE sample. Some of the possible duplicated Cepheids were removed in the OGLE database by consulting table 4 of \citet{uda99b}.

     \item Cepheids with unusual colour. We first plot out (as in Figure \ref{figcut}[a]) the extinction corrected PC relation at mean light. The plot shows that there are number of outliers in the period-colour plane, mostly with $\log(P)<1.0$. The presence of these outliers is probably due to: (a) their extinction is either over- or under-estimated; (b) they have blue or red companions that cannot be resolved due to the problems of blending; or (c) other unknown physical reasons. A detailed investigation of these outliers is beyond the scope of this paper, but it is clear that they should be removed from the sample. These outliers are removed with the adopted colour-cut of $0.35<(V-I)^{\mathrm{mean}}_0<0.95$, a compromise between maximizing the number of Cepheids in the sample and excluding the Cepheids with unusual colour\footnote{Note that this colour-cut is only applied to the short period Cepheids because there are not many long period Cepheids in the sample. Furthermore, the PC (mean) relation for the long period Cepheids is steeper than the short period Cepheids, hence as the period increases the colour becomes redder, which can be excluded from the colour-cut.}.

     \item Cepheids with unusually low (or high) amplitude. Some Cepheids with unusually low $V$- and $I$-band amplitudes were found in the sample. Their amplitudes are typically $2\sim3$ times smaller as compared to the amplitudes of other Cepheids at given period. Some examples of the light curves for these low amplitude Cepheids are given in \citet{kan03}. In addition, most of the light curves for these low amplitude Cepheids can be fitted with $n=4$ Fourier expansion, while other Cepheids with ``normal'' amplitude may require higher order fits. \citet{kan03} has briefly discussed some possible physical reasons for these Cepheids to have such low amplitudes, e.g. they are just entering or leaving the fundamental mode instability strip \citep{buc02} or they have different chemical composition \citep[see, e.g.,][]{pac00}. The detailed investigation of these low amplitudes Cepheids is beyond the scope of this paper. Here, we apply a conservative amplitude cut of $0.3$ mag. in the $V$-band to remove the low amplitude Cepheids. Besides that, we also remove OGLE-286532 (with unusually low amplitude) and HV-2883 (with unusually high amplitude) as they are clear outliers in the $\log(P)$-amplitude plot \citep[not shown, but see][]{nge05b}. Note that \citet{pie04} applied a cut of $0.4$ mag. to remove the low amplitude Cepheids in NGC 6822. Other examples of removing the low-amplitude Cepheids can also be found in \citet{way84}.
     
     \item Cepheids with $\log(P)<0.4$ and $\log(P)>1.8$. In order to guard against possible contamination from the first overtone Cepheids \citep{uda99} and to be consistent with the previous studies \citep{kan04,san04,uda99}, we removed Cepheids with $\log(P)<0.4$ (see further justification in \citealt{nge05}). Regarding the removal of Cepheids with $\log(P)>1.8$, a preliminary analysis of the PC relation reveals that few of the longest Cepheids should be removed from the sample, because they are clear outliers in the PC plot at maximum light (see upper panel of Figure \ref{figpcmax}). Without these longest period Cepheids, the PC(max) relation for the long period Cepheids is flat, which is consistent with the results found in Paper I. The hypothesis of the HIF-photosphere interaction also suggests the flatness of the PC relation at maximum light for long period Cepheids. However, as the period gets longer (with $\log[P]>1.5$), the photosphere disengages from the HIF \citep{sim93}. These longest period Cepheids have biased the slope of the PC(max) relation by making the slope becomes steeper.
     \end{enumerate}

%**********************************************************
%      FIGURE: selection of good Cepheid
%**********************************************************
 
    \ni These selection criteria are guided mainly by the philosophy that it is better to lose some ``bad'' but real Cepheids rather than including those spurious and doubtful Cepheids in the sample \citep{leo03}, or those with bad fitted light curves that will give inaccurate measurements of the maximum and minimum light. Hence, the final sample consists of 641 LMC Cepheids that will be considered further. The locations of the outliers from various selection criteria are shown in Figure \ref{figcut} for the PC(mean) relation, $V$-band PL relation, $R_{21}$-$\log(P)$ relation\footnote{$R_{21}=A_2/A_1$, where $A_n$ are the Fourier amplitudes. See \citet{sim81} and \citet{nge03} for details.} and the colour-magnitude diagram (CMD). Note that some of the outliers are located within the ``good'' Cepheids. However they can be eliminated due to various physical reasons as given above, especially those with poorly fit light curves that will give inaccurate measurements at maximum, mean and minimum light. A simple sigma-clipping algorithm \citep[e.g.,][]{uda99} will not be able to remove these outliers \citep{kan03}.

\section{The New Empirical PC and AC Relations}

      To construct the empirical PC \& AC relations, we used the following quantities from the Fourier fits to the Cepheid data as obtained from previous section:

     \begin{itemize}
     \item $V$-band amplitude: the difference of the numerical maximum and minimum from the Fourier expansion, $V_{amp}=V_{min}-V_{max}$.
     \item $(V-I)^{max}$: defined as $V_{max}-I_{phmax}$, where $I_{phmax}$ is the $I$-band magnitude at the same phase as $V_{max}$.
     \item $(V-I)^{mean}$: defined as $A_0(V)-A_0(I)$, where $A_0$ is the mean value from the Fourier expansion \citep[see][]{nge03}. This is very similar to the conventional definition of the mean colour, $<V>-<I>$, where $<>$ denotes the intensity mean.
     \item $(V-I)^{phmean}$: defined as $V_{mean}-I_{phmean}$, where $I_{phmean}$ is the $I$-band magnitude at the same phase as $V_{mean}$. $V_{mean}$ is the $V$-band magnitude closest to $A_0(V)$, the mean value from Fourier expansion.
     \item $(V-I)^{min}$: defined as $V_{min}-I_{phmin}$, where $I_{phmin}$ is the $I$-band magnitude at the same phase as $V_{min}$.
     \end{itemize}

\begin{figure*}
         \vspace{0cm}
         \hbox{\hspace{0.2cm}\epsfxsize=7.5cm \epsfbox{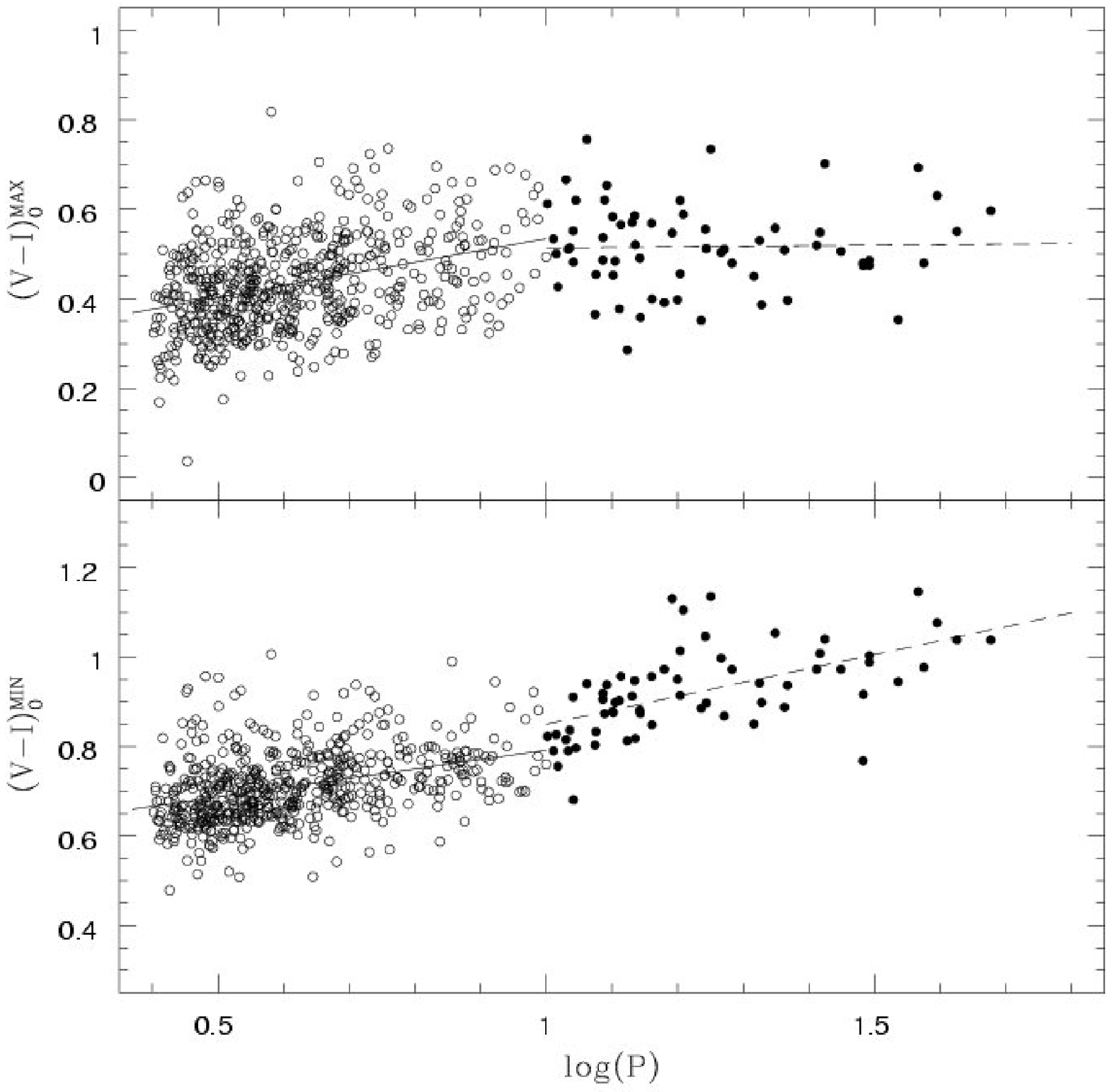}
           \epsfxsize=7.5cm \epsfbox{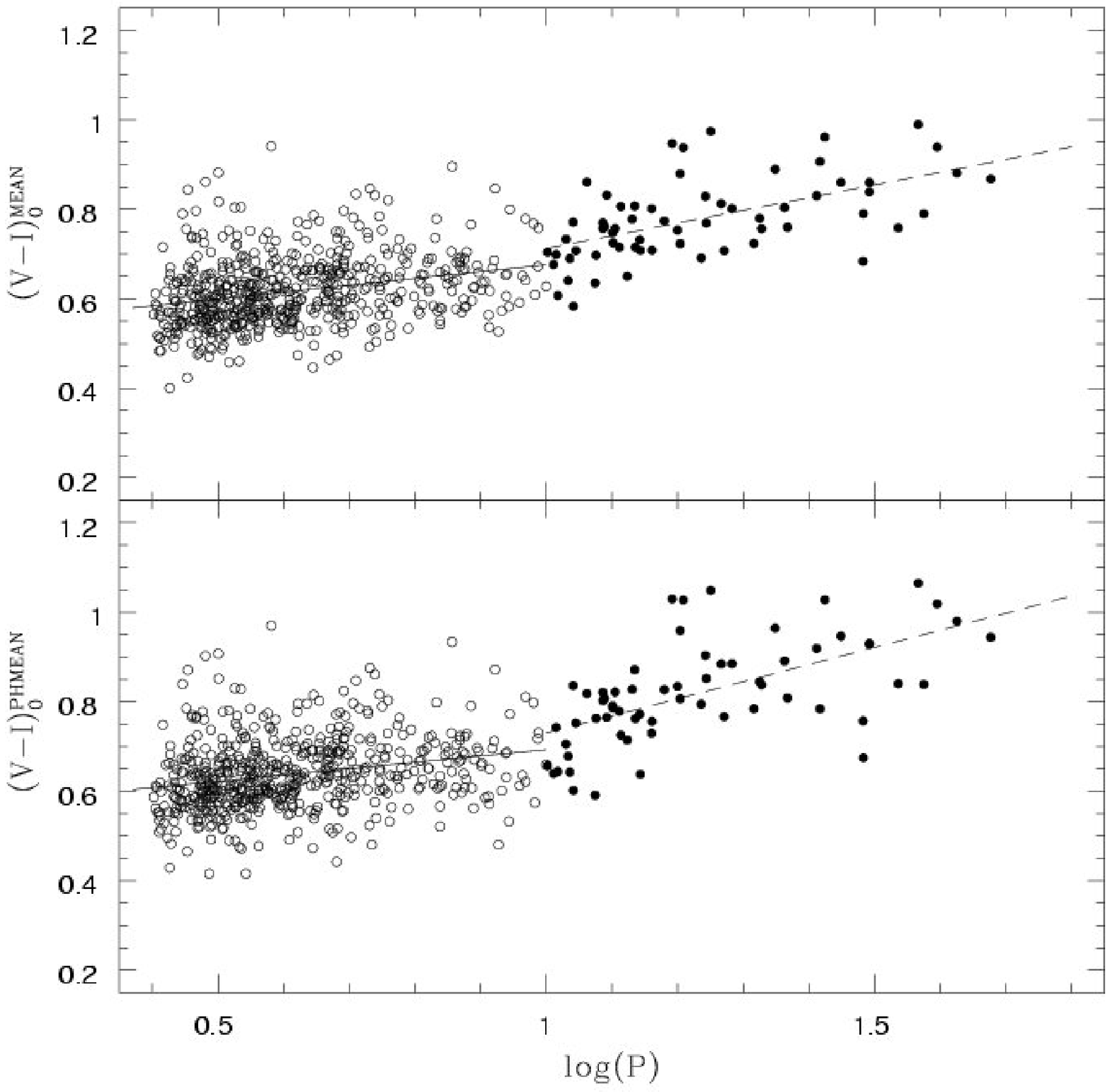}}
         \vspace{0cm}
         \caption{The period-colour (PC) relations for the LMC Cepheids at maximum, means and minimum light. The open and filled circles are for short and long period Cepheids, respectively. The solid and dashed lines are the fitted PC relations for the short and long period Cepheids, respectively. \label{c9figpc}}
       \end{figure*}

\begin{figure*}
         \vspace{0cm}
         \hbox{\hspace{0.2cm}\epsfxsize=7.5cm \epsfbox{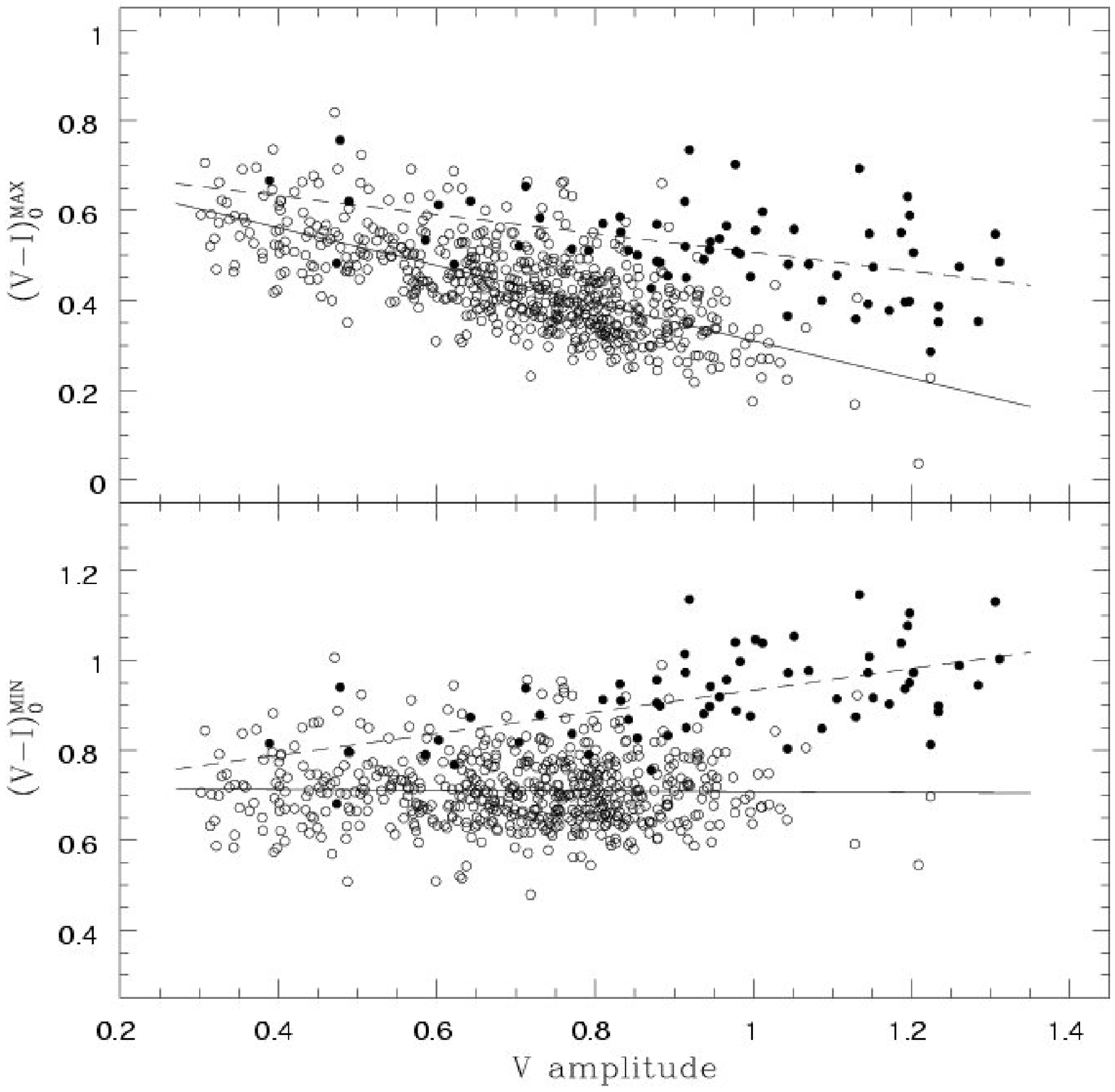}
           \epsfxsize=7.5cm \epsfbox{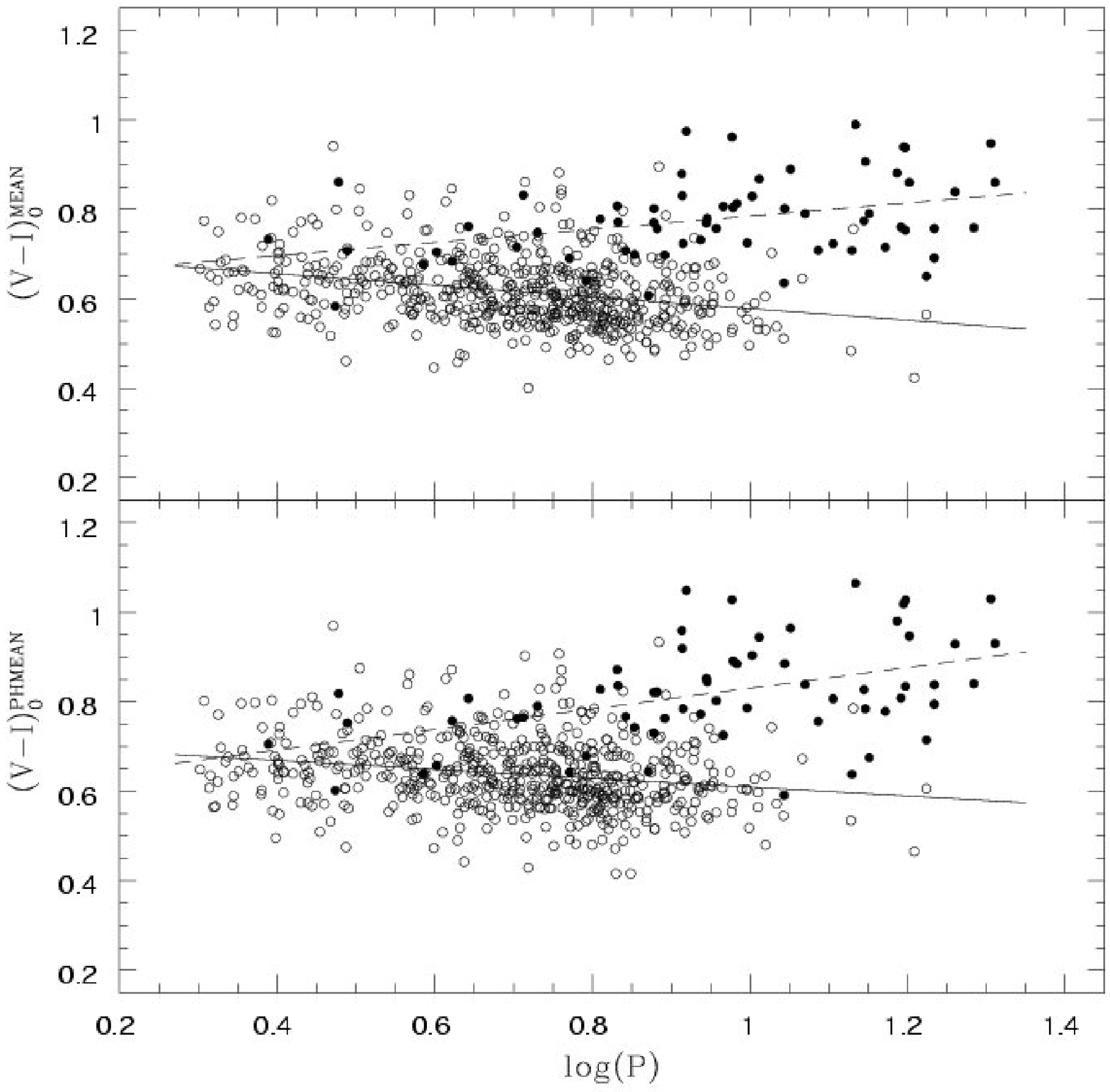}}
         \vspace{0cm}
         \caption{Same as Figure \ref{c9figpc}, but for the amplitude-colour (AC) relations. \label{c9figac}}
       \end{figure*}

     \ni  These quantities have been corrected for extinction as mentioned in previous section. The empirical LMC PC and AC relations at maximum, mean and minimum light for all, long and short period Cepheids are summarized in Table \ref{c9tabpc} \& \ref{c9tabac}, and the corresponding plots are presented in Figure \ref{c9figpc} \& \ref{c9figac}, respectively. 

       \begin{table}
         \centering
         \caption{The LMC period-colour relation in the form of $(V-I)=a\log(P)+b$, and $\sigma$ is the dispersion of the relation.}
         \label{c9tabpc}
         \begin{tabular}{lccc} \hline
           Phase & $a$ & $b$ & $\sigma$ \\
           \hline 
           \multicolumn{4}{c}{All, $N=641$} \\   
           Maximum & $0.168\pm0.017$ & $0.328\pm0.012$ & 0.099 \\
           Mean    & $0.228\pm0.013$ & $0.479\pm0.009$ & 0.075 \\
           Phmean  & $0.248\pm0.014$ & $0.489\pm0.010$ & 0.081 \\
           Minimum & $0.297\pm0.013$ & $0.532\pm0.009$ & 0.075 \\
           \multicolumn{4}{c}{Long, $N=63$} \\   
           Maximum & $0.012\pm0.069$ & $0.502\pm0.086$ & 0.098 \\
           Mean    & $0.284\pm0.053$ & $0.428\pm0.066$ & 0.075 \\
           Phmean  & $0.383\pm0.065$ & $0.347\pm0.081$ & 0.092 \\
           Minimum & $0.311\pm0.056$ & $0.539\pm0.070$ & 0.081 \\
           \multicolumn{4}{c}{Short, $N=578$} \\   
           Maximum & $0.263\pm0.029$ & $0.272\pm0.018$ & 0.097 \\
           Mean    & $0.153\pm0.022$ & $0.523\pm0.014$ & 0.074 \\
           Phmean  & $0.141\pm0.023$ & $0.552\pm0.014$ & 0.078 \\
           Minimum & $0.210\pm0.022$ & $0.582\pm0.014$ & 0.073 \\
           \hline
         \end{tabular}
       \end{table}

       \begin{table}
         \centering
         \caption{The LMC amplitude-colour relation in the form of $(V-I)=aV_{amp}+b$, and $\sigma$ is the dispersion of the relation.}
         \label{c9tabac}
         \begin{tabular}{lccc} \hline
           Phase & $a$ & $b$ & $\sigma$ \\
           \hline  
           \multicolumn{4}{c}{All, $N=641$} \\   
           Maximum & $-0.274\pm0.019$ & $0.641\pm0.015$ & 0.093 \\
           Mean    & $ 0.029\pm0.019$ & $0.611\pm0.014$ & 0.092 \\
           Phmean  & $ 0.074\pm0.021$ & $0.602\pm0.016$ & 0.099 \\
           Minimum & $ 0.159\pm0.020$ & $0.616\pm0.015$ & 0.098 \\
           \multicolumn{4}{c}{Long, $N=63$} \\   
           Maximum & $-0.209\pm0.049$ & $0.716\pm0.048$ & 0.086 \\
           Mean    & $ 0.148\pm0.048$ & $0.637\pm0.047$ & 0.085 \\
           Phmean  & $ 0.230\pm0.058$ & $0.600\pm0.057$ & 0.103 \\
           Minimum & $ 0.240\pm0.046$ & $0.693\pm0.045$ & 0.082 \\
           \multicolumn{4}{c}{Short, $N=578$} \\   
           Maximum & $-0.419\pm0.019$ & $0.729\pm0.014$ & 0.076 \\ 
           Mean    & $-0.130\pm0.018$ & $0.708\pm0.013$ & 0.074 \\
           Phmean  & $-0.099\pm0.109$ & $0.708\pm0.014$ & 0.078 \\
           Minimum & $-0.008\pm0.020$ & $0.716\pm0.014$ & 0.079 \\
           \hline
         \end{tabular}
       \end{table}

     To test the non-linearity of the PC and AC relations, or the ``break'' at a period of 10 days, we apply the $F$-test as given in Paper I and in \citet{nge05}. The null hypothesis in the $F$-test is single line regression is sufficient, while the alternate hypothesis is that two lines regressions with a discontinuity (a break) at 10 days is necessary to fit the data. The probability $p(F)$, under the null hypothesis, can be obtained with the corresponding $F$-values and the degrees of freedom. In general, the large value of $F$ (equivalent to the small value of $p[F]$) indicates that the null hypothesis can be rejected. For our sample, $F\sim3.0$ when $p(F)=0.05$ (the 95\% confident level), therefore the null hypothesis can be rejected if the $F$-value is greater than $3$ with more than 95\% confident level and the data is more consistent with the two-line regression. A glance of Table \ref{c9tabpc} and Figure \ref{c9figpc} suggests that the LMC PC relations are broken at maximum, mean and minimum light. These are confirmed with the $F$-test results with $F_{PC}(\mathrm{max,mean,phmean,min})=\{9.49,\ 8.59,\ 16.8,\ 11.8\}$. Similarly, the $F$-test results for the AC relation are: $F_{AC}(\mathrm{max,mean,phmean,min})=\{139,\ 162,\ 157,\ 171\}$. Hence, the LMC PC and AC relations are non-linear (hence broken) at maximum, means and minimum light. Note that the flatness of the long period PC(max) relation as given in Table \ref{c9tabpc} ($0.012\pm0.069$) is in good agreement with the slope found in Paper I ($-0.031\pm0.101$). Recall that equation (1) predicts that if the PC relation is flat at maximum light, then there is a correlation between the amplitude and the colour at minimum light. This is seen in Table \ref{c9tabac} (and in Figure \ref{c9figac}) for the long period AC(min) relation, with a slope of $0.240\pm0.046$. 

%**********************************************************
%      FIGURE: empirical PC relations
%**********************************************************
 
%**********************************************************
%      FIGURE: empirical AC relations
%**********************************************************

\section{Methods and LMC Models}

     The stellar pulsation codes we used are both linear \citep{yec98} and non-linear \citep{kol02}.  These codes, which include a 1-D turbulent convection recipe \citep{yec98}, are the same as in Paper II. Briefly speaking, the codes take the mass ($M$), luminosity ($L$), effective temperature ($T_{eff}$) and chemical composition ($X,Z$) as input parameters. The chemical composition is set to be $(X,Z)=(0.70,0.008)$ to represent the LMC hydrogen and metallicity abundance (by mass). The mass and luminosity are obtained from the ML relations calculated from evolutionary models. The $T_{eff}$ are chosen to ensure the models oscillate in the fundamental mode and located inside the Cepheid instability strip. The pulsation periods for the models are obtained from a linear non-adiabatic analysis \citep{yec98}. All other parameters used in the pulsation codes had the same values for the LMC and Galactic models (Paper II). This included the $\alpha$ parameters that are part of the turbulent convection recipe, though see Section 8. Of course, one variable parameter was the metallicity. The only other difference between this study and Paper II, besides the metallicity, is the value set for the artificial viscosity parameter, $C_q$. In this study, we set $C_q=16.0$ for the LMC models to improve the shape of the theoretical light curves, in contrast to the value of $4.0$ used for the Galactic models.   

     In Paper II, the ML relations are adopted from \citet{chi89} and \citet{bon00}. In order to be consistent with previous work, the ML relations used in this paper will also be adopted from these two sources. However, \citet{chi89} only provided two ML relations, one for $Z=0.020$ which are used in Paper II, and another one for $Z=0.001$. Hence we have to adopt the second ML relation for the LMC models. Even though the LMC metallicity is higher than $Z=0.001$, the LMC is still considered as a low metallicity system in the literature. Hence the \citet{chi89} ML relation can be approximately applied for the LMC models. An anonymous referee pointed out that an interpolation of the \citet{chi89} ML relations between $Z=0.020$ and $Z=0.001$ should also be used. We have included the interpolated ML relation in our model calculations. In the context of the HIF-photosphere interaction, it is the ML relation which dictates at what period and at what phases this will occur. Stellar evolutionary theory changes the ML relation as a function of metallicity. Hence the coefficients of the ML relation are important in determining the nature of the HIF-photosphere interaction (Paper II). In short, the ML relations used are:

     \begin{table}
       \centering
       \caption{Input parameters for LMC Cepheid models with periods obtained from a linear analysis. The periods, $P_0$ and $P_1$, are referred to the fundamental and first overtone periods, respectively. Similarly for the growth rate, $\eta$. Both of the mass and luminosity are in Solar units, the temperature is in K and the period is in days.}
       \label{tabinput}
       \begin{tabular}{ccccccc} \hline
         $M$ & $\log(L)$ & $T_{eff} $ & $P_0$ & $\eta_0$ & $P_1$ & $\eta_1$ \\
         \hline 
         \multicolumn{7}{c}{ML Relation from \citet{bon00}} \\
         11.0 & 4.375 & 5050 & 46.4155 & 0.124 & 28.98 & -0.118 \\
         10.0 & 4.236 & 5100 & 35.6727 & 0.091 & 22.92 & -0.093 \\
         9.50 & 4.161 & 5250 & 28.2406 & 0.094 & 18.68 & -0.046 \\
         9.10 & 4.099 & 5260 & 25.3960 & 0.082 & 16.92 & -0.042 \\
         8.75 & 4.042 & 5310 & 22.3804 & 0.076 & 15.07 & -0.027 \\
         8.40 & 3.982 & 5380 & 19.3886 & 0.071 & 13.20 & -0.008 \\
         7.95 & 3.902 & 5330 & 17.7750 & 0.055 & 12.09 & -0.027 \\
         7.00 & 3.717 & 5410 & 12.6085 & 0.035 & 8.722 & -0.018 \\
         6.55 & 3.620 & 5490 & 10.2940 & 0.031 & 7.183 & -0.006 \\
         6.40 & 3.587 & 5485 & 9.81474 & 0.027 & 6.853 & -0.010 \\
         6.00 & 3.493 & 5510 & 8.37226 & 0.020 & 5.866 & -0.014 \\
         5.90 & 3.468 & 5500 & 8.12498 & 0.017 & 5.691 & -0.018 \\
         5.80 & 3.443 & 5525 & 7.69466 & 0.017 & 5.400 & -0.015 \\
         5.70 & 3.418 & 5560 & 7.23505 & 0.017 & 5.090 & -0.009 \\
         5.30 & 3.312 & 5600 & 6.01283 & 0.012 & 4.244 & -0.009 \\ 
         \multicolumn{7}{c}{ML Relation from \citet{chi89}} \\
         7.20 & 4.272 & 5380 & 40.2561 & 0.275 & 24.51 & -0.162 \\
         6.80 & 4.192 & 5380 & 35.4374 & 0.264 & 21.91 & -0.122 \\
         6.20 & 4.063 & 5410 & 28.2629 & 0.225 & 17.94 & -0.076 \\
         5.95 & 4.005 & 5420 & 25.6378 & 0.211 & 16.43 & -0.060 \\
         5.40 & 3.869 & 5510 & 19.4314 & 0.170 & 12.82 & -0.015 \\
         5.15 & 3.803 & 5510 & 17.5523 & 0.160 & 11.66 & -0.007 \\
         4.65 & 3.660 & 5490 & 14.3143 & 0.131 & 9.611 & -0.010 \\
         4.20 & 3.518 & 5510 & 11.3659 & 0.101 & 7.729 & -0.011 \\
         4.00 & 3.450 & 5545 & 10.0011 & 0.089 & 6.854 & -0.005 \\
         3.95 & 3.432 & 5540 & 9.77393 & 0.085 & 6.701 & -0.008 \\
         3.80 & 3.378 & 5550 & 8.94637 & 0.075 & 6.157 & -0.009 \\
         3.70 & 3.341 & 5575 & 8.31297 & 0.070 & 5.745 & -0.005 \\
         3.65 & 3.322 & 5570 & 8.10751 & 0.066 & 5.605 & -0.008 \\
         3.60 & 3.302 & 5530 & 8.09994 & 0.058 & 5.583 & -0.022 \\
         3.60 & 3.302 & 5600 & 7.71463 & 0.065 & 5.352 & -0.001 \\
	 \multicolumn{7}{c}{ML Relation from interpolated \citet{chi89}} \\
	 6.80 & 4.092 & 5280 & 30.9661 & 0.207 & 19.51 & -0.091 \\
	 5.20 & 3.701 & 5340 & 15.9744 & 0.094 & 10.63 & -0.050 \\
	 4.40 & 3.457 & 5460 & 10.0562 & 0.056 & 6.898 & -0.028 \\
	 4.20 & 3.389 & 5550 & 8.51317 & 0.054 & 5.906 & -0.007 \\
	 3.80 & 3.243 & 5630 & 6.46303 & 0.039 & 4.532 & -0.002 \\
         \hline 
       \end{tabular}
     \end{table}

     \begin{enumerate}   
     \item ML relation given in \citet{bon00}:
       \begin{eqnarray}
         \log(L) & = & 0.90 + 3.35\log(M) + 1.36\log(Y) - 0.34\log(Z), \nonumber \\
         & = & 3.35\log(M) + 0.886.
       \end{eqnarray}
     
     \item ML relation given in \citet{chi89}:
       \begin{eqnarray}
         \log(L) & = & 3.22\log(M)+1.511.
       \end{eqnarray}

     \item ML relation interpolated between two \citet{chi89} relations at $Z=0.02$ and $Z=0.001$ to yield a relation at $Z=0.008$:
       \begin{eqnarray}
         \log(L) & = & 3.36\log(M)+1.295.
       \end{eqnarray}
     \end{enumerate}
     
     \ni The units for both $M$ and $L$ are in Solar units. Note that these ML relations cover reasonably broad $L/M$ ratios given in the literature. The input parameters for the LMC models with these ML relations and the periods calculated from linear non-adiabatic analysis are given in Table \ref{tabinput}. 

     After the full amplitude models are constructed from the pulsation codes, the temperature and the opacity profile can be plotted in terms of the internal mass distribution ($\log[1-M_r/M]$, where $M_r$ is mass within radius $r$ and $M$ is the total mass) at a given phase of pulsation. As in Paper II, the locations of the HIF (sharp rise in the temperature profile) and photosphere (at optical depth $\tau=2/3$) can be identified in the temperature profile. To quantify the HIF-photosphere interaction (if the photosphere is next to the base of the HIF or not, see also Paper II), we calculate the ``distance'', $\Delta$, in $\log(1-M_r/M)$ between the HIF and the photosphere from the temperature profile. The definition of $\Delta$ can be found in Paper II. A small $\Delta$ means there is a HIF-photosphere interaction, and vice versa. 

     The theoretical quantities from the models can be compared to the observed quantities using the following prescriptions:

     \begin{table}
       \centering
       \caption{Temperatures at maximum and minimum light from full-amplitude non-linear model calculations. The periods, luminosity and temperature are in days, $L_{\odot}$ and K, respectively. \label{c9tabmaxmin}}
       \begin{tabular}{ccccc} \hline
         $P$ & $L_{max}$ & $T_{max}$ & $L_{min}$ & $T_{min}$ \\ 
         \hline 
         \multicolumn{5}{c}{ML Relation from \citet{bon00}} \\ 
         46.4155 & 27481.28 & 5445.00 & 18329.74 & 4826.00 \\
         35.6727 & 19742.83 & 5434.01 & 14321.20 & 4962.65 \\
         28.2406 & 16894.15 & 5502.35 & 12093.99 & 4978.72 \\
         25.3960 & 14460.62 & 5562.28 & 10570.70 & 5010.85 \\
         22.3804 & 12772.75 & 5615.34 & 9283.257 & 5069.43 \\
         19.3886 & 11263.84 & 5687.98 & 8143.460 & 5145.20 \\
         17.7750 & 9034.392 & 5561.39 & 7044.743 & 5147.04 \\
         12.6085 & 5637.467 & 5528.15 & 4800.840 & 5278.11 \\
         10.2940 & 4371.170 & 5537.42 & 3839.001 & 5377.80 \\
         9.81474 & 4004.569 & 5518.39 & 3580.396 & 5383.61 \\
         8.37226 & 3217.688 & 5596.34 & 2929.354 & 5437.17 \\
         8.12498 & 3032.639 & 5581.53 & 2793.441 & 5438.96 \\
         7.69466 & 2864.745 & 5602.37 & 2640.439 & 5467.54 \\
         7.23505 & 2705.430 & 5639.03 & 2490.271 & 5504.98 \\
         6.01283 & 2105.589 & 5669.57 & 1986.586 & 5567.91 \\ 
         \multicolumn{5}{c}{ML Relation from \citet{chi89}} \\ 
         40.2561 & 23500.98 & 5821.99 & 11853.38 & 4947.11 \\
         35.4374 & 19604.60 & 5830.18 & 9898.124 & 4962.26 \\
         28.2629 & 14661.29 & 5875.98 & 7659.896 & 5057.96 \\
         25.6378 & 12836.04 & 5884.01 & 6916.299 & 5108.02 \\
         19.4314 & 9538.226 & 5985.21 & 5555.179 & 5303.60 \\
         17.5523 & 8035.700 & 5921.07 & 4947.967 & 5142.58 \\
         14.3143 & 5490.345 & 5783.01 & 3639.964 & 5164.11 \\
         11.3659 & 3850.971 & 5814.49 & 2722.473 & 5242.47 \\
         10.0011 & 3286.030 & 5817.00 & 2377.950 & 5300.52 \\
         9.77393 & 3136.139 & 5801.77 & 2301.751 & 5296.73 \\
         8.94637 & 2738.003 & 5798.30 & 2071.106 & 5331.19 \\
         8.31297 & 2499.416 & 5794.49 & 1917.696 & 5365.34 \\
         8.10751 & 2374.713 & 5780.35 & 1850.048 & 5369.40 \\
         8.09994 & 2235.292 & 5723.18 & 1796.586 & 5355.41 \\
         7.71463 & 2271.393 & 5805.49 & 1772.270 & 5408.34 \\ 
	 \multicolumn{5}{c}{ML Relation from interpolated \citet{chi89} } \\
         30.9661 & 14958.03 & 5611.77 & 8486.049 & 4963.81 \\
	 15.9744 & 5697.042 & 5665.97 & 4209.487 & 5079.81 \\
	 10.0562 & 3197.451 & 5670.64 & 2560.871 & 5283.85 \\
	 8.51317 & 2723.566 & 5716.48 & 2201.628 & 5382.21 \\
         6.46303 & 1852.537 & 5684.63 & 1593.714 & 5507.38 \\
	 \hline 
       \end{tabular}
     \end{table}

%************************************************
%  FIGURE: atmosphere compare
%************************************************

     \begin{figure}
       \centering 
       \epsfxsize=7.5cm{\epsfbox{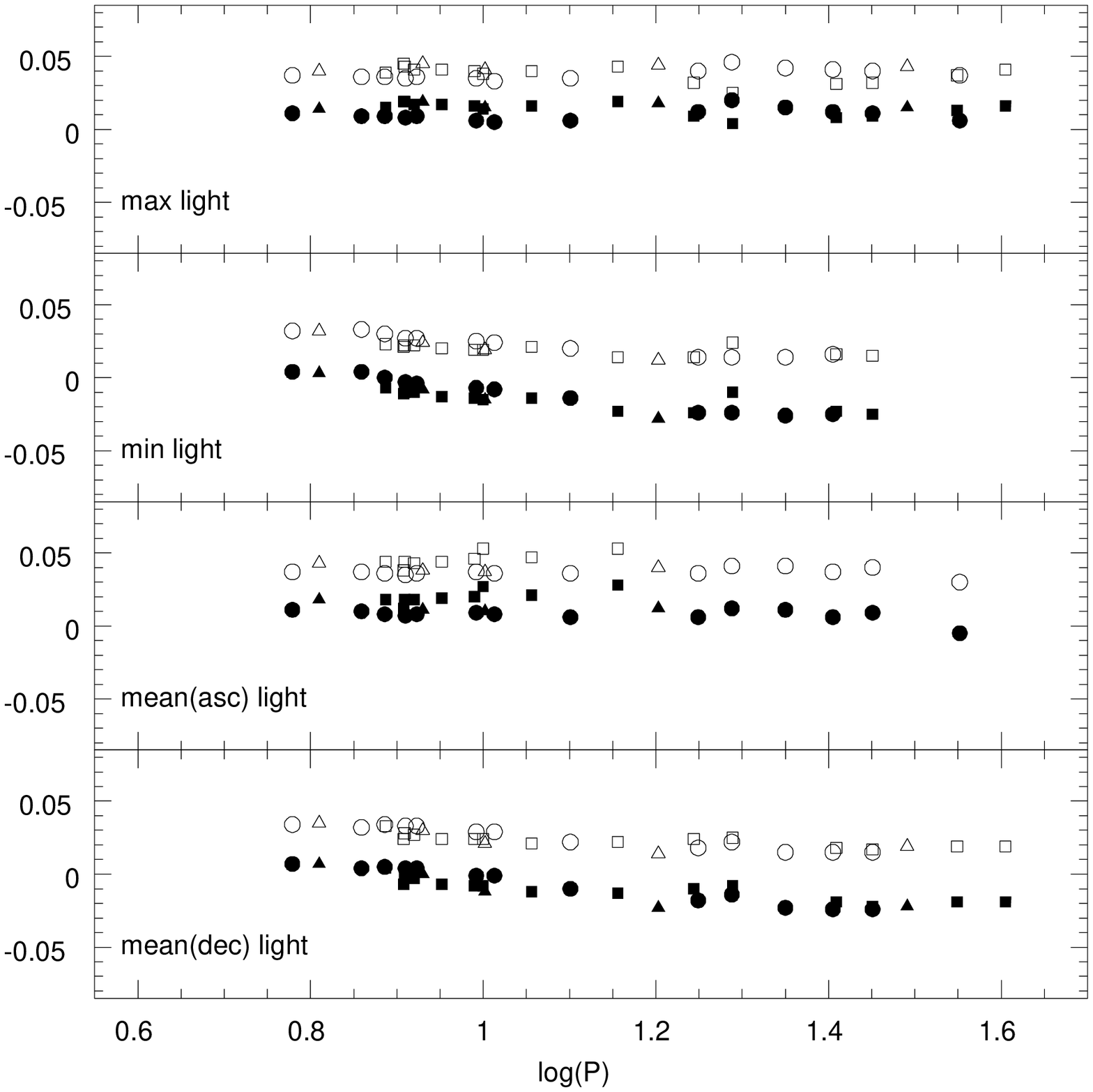}}
       \caption{Comparison of the colours at maximum, mean and minimum light from the models using different atmosphere databases. The $y$-axis indicates the difference when using the {\tt BaSeL} atmosphere and the {\tt SBT} atmosphere, i.e. $BaSeL-SBT$. The circles, squares and triangles are for the models calculated with \citet{bon00}, \citet{chi89} and interpolated \citet{chi89} ML relations, respectively. Open and filled symbols correspond to the use of micro-turbulence velocity of $1.7\mathrm{kms}^{-1}$ and $5.0\mathrm{kms}^{-1}$, respectively, from the {\tt SBT} atmosphere database.}
       \label{figatmos}
     \end{figure}

     \begin{enumerate}
     \item  As in Paper II, we use the {\tt BaSeL} atmosphere database \citep{lej02,wes02} to construct a fit giving temperature and effective gravity as a function of $(V-I)$ colour. The effective gravity is obtained at the appropriate phase from the models (see Paper II). These prescriptions are used to convert the temperatures to the $(V-I)$ colours. The bolometric corrections ($BC$) are obtained in a similar manner. The anonymous referee has suggested that $(V-I)$ may not be a good way to convert between temperature and colour unless both of the micro-turbulence and surface gravity are included. As indicated above this is the case, and in any case our results and those of Paper II for Galactic models, show good agreement between theory and observations. A number of previous authors have used this method and some authors commented that this colour can be used as an indicator of temperature \citep[e.g.][]{bea01,tam03}. The empirical relations we studied in this series were also mainly in the $(V-I)$ colour.

     \item  In addition to the {\tt BaSeL} atmosphere, we also use the atmosphere fit from \citet{san99}, referring this as the {\tt SBT} atmosphere in our paper. The {\tt SBT} atmosphere does include both of the effective gravity and the micro-turbulence in their table 6 for the temperature and colour conversion. These conversions are tabulated for two micro-turbulence velocities of $1.7\mathrm{kms}^{-1}$ and $5.0\mathrm{kms}^{-1}$, as well as for various metallicities. To apply these conversions to our LMC models, we first interpolated the conversions between $[A/H]=0.0$ and $[A/H]=-0.5$ to $[A/H]=-0.3$, which is appropriate for the LMC metallicity. The $(V-I)$ colours at the maximum, mean and minimum light are then obtained from the given effective temperature and the effective gravity for both of the micro-turbulence velocities. 

     \item  We use the prescriptions given in \citet{bea01} to convert the observed colours to the temperatures appropriate for the LMC data as follows:
       \begin{eqnarray}
         \log(g) & = & 2.62 - 1.21 \log (P), \nonumber \\
         \log(T_{eff}) & = & 3.91545 + 0.0056\log(g) - 0.2487(V-I)_0, \nonumber \\
         \Delta T & = & \log(T_{eff}) - 3.772, \nonumber \\
         BC & = & -0.0153 + 2.122\Delta T - 0.02\log(g)  \nonumber \\ 
            &   & - 11.65(\Delta T)^2.\nonumber 
       \end{eqnarray}

       \ni Note that these functions are also obtained from the {\tt BaSeL} atmosphere database. 
     \end{enumerate}

     We can compare the colours obtained from the {\tt BaSeL} and {\tt SBT} atmosphere for our models constructed in this paper. The results are presented in Figure \ref{figatmos}. From this figure it can be seen that the colours obtained from both of the atmosphere fits agree within $0.05$mag. level. The difference is even smaller if the micro-turbulence velocity of $5.0\mathrm{kms}^{-1}$ is used. This indicates that the $(V-I)$ colours can be used to indicate the temperature. Since the results of our models are qualitatively compared to the observations (see next section) and not used to quantitatively derive any theoretical PC and/or AC relations, an accuracy of $\sim0.05$mag., independent of period, from the atmosphere fit is acceptable and would not cause problems for our results. Note that the {\tt SBT} atmosphere are only defined for $5000K\leq T_{eff}\leq 7000K$ and $0.75\leq \log(g)\leq 3.00$, few of our models either the $T_{eff}$ or $\log(g)$ or both are beyond these ranges at certain phases, hence no colours can be obtained from the {\tt SBT} atmosphere (for example some points are missing at minimum light for few of the long period models, as shown in Figure \ref{figatmos}). Due to these reasons, we continue adopt the {\tt BaSeL} atmosphere fits to convert the temperatures and $(V-I)$ colours, after taking account of the effective gravity in the fits, as a function of phase.

\section{Results from the Models}

     \begin{table*}
       \centering
       \caption{Temperatures at mean light from full-amplitude non-linear model calculations. See Paper II for the meanings of $<L>$, $L_{mean}$, $T_{mean}$ and $T^{inter}_{mean}$. The periods, luminosity and temperature are in days, $L_{\odot}$ and K, respectively.}
       \label{c9tabmean}
       {\footnotesize
       \begin{tabular}{cccccccc} \hline
         $P$ & $<L>$ & $L_{mean}(asc)$ & $T_{mean} (asc)$ & $L_{mean} (des)$ & $T_{mean} (des)$ & $T_{mean}^{inter}$ (asc) & $T_{mean}^{inter}$ (des)\\ 
         \hline 
         \multicolumn{8}{c}{ML Relation from \citet{bon00}} \\
         46.4155 & 24249.2 & 24423.695 & 5330.06 & 24282.129 & 4882.55 & 5319.42 & 4880.66 \\
         35.6727 & 17207.7 & 17078.471 & 5293.90 & 17188.051 & 4922.99 & 5305.14 & 4924.47 \\
         28.2406 & 14484.7 & 14466.869 & 5457.79 & 14529.348 & 5073.65 & 5459.55 & 5069.64 \\
         25.3960 & 12540.8 & 12461.087 & 5443.66 & 12544.735 & 5096.58 & 5452.77 & 5096.18 \\
         22.3804 & 10997.1 & 10996.587 & 5493.44 & 11039.123 & 5157.95 & 5493.51 & 5152.87 \\
         19.3886 & 9587.02 & 9560.1922 & 5545.87 & 9618.0583 & 5232.85 & 5549.58 & 5228.51 \\
         17.7750 & 7983.35 & 7988.5457 & 5473.94 & 7994.9159 & 5213.83 & 5473.06 & 5211.77 \\
         12.6085 & 5208.60 & 5208.1564 & 5493.34 & 5202.6440 & 5349.00 & 5493.36 & 5350.84 \\
         10.2940 & 4170.31 & 4162.1135 & 5560.81 & 4174.0993 & 5450.34 & 5564.43 & 5449.06 \\
         9.81474 & 3860.04 & 3861.4313 & 5556.01 & 3852.8715 & 5446.51 & 5555.34 & 5449.11 \\
         8.37226 & 3109.66 & 3106.7337 & 5570.85 & 3110.1806 & 5481.59 & 5572.55 & 5481.37 \\
         8.12498 & 2939.41 & 2935.0599 & 5553.67 & 2939.6504 & 5476.34 & 5556.38 & 5476.23 \\
         7.69466 & 2775.73 & 2775.1245 & 5581.66 & 2777.4057 & 5501.67 & 5582.05 & 5500.77 \\
         7.23505 & 2618.53 & 2620.1965 & 5620.29 & 2618.8674 & 5534.39 & 5619.15 & 5534.20 \\
         6.01283 & 2051.97 & 2050.8394 & 5646.29 & 2051.3277 & 5583.97 & 5647.27 & 5584.50 \\
         \multicolumn{8}{c}{ML Relation from \citet{chi89}} \\  
         40.2561 & 18754.3 & 18580.019 & 5704.60 & 18824.542 & 5144.86 & 5718.84 & 5141.42 \\
         35.4374 & 15896.6 & 15885.248 & 5749.07 & 15861.179 & 5147.12 & 5750.16 & 5158.63 \\
         28.2627 & 11157.8 & 11188.594 & 5692.54 & 11143.155 & 5111.26 & 5688.21 & 5112.18 \\
         25.6378 & 10388.8 & 10366.128 & 5782.33 & 10358.696 & 5184.81 & 5785.59 & 5188.79 \\
         19.4314 & 7743.89 & 7741.3087 & 5871.14 & 7750.7288 & 5332.20 & 5871.60 & 5330.76 \\
         17.5523 & 6538.19 & 6537.3819 & 5833.02 & 6532.0823 & 5310.84 & 5833.20 & 5312.28 \\
         14.3143 & 4564.83 & 4621.4163 & 5750.64 & 4565.7876 & 5280.59 & 5732.80 & 5280.34 \\
         11.3659 & 3289.17 & 3288.4584 & 5711.02 & 3281.3799 & 5328.36 & 5711.31 & 5331.32 \\
         10.0011 & 2811.35 & 2823.6619 & 5726.94 & 2800.7972 & 5376.66 & 5721.20 & 5381.50 \\
         9.77393 & 2702.54 & 2712.0209 & 5715.81 & 2692.4263 & 5378.63 & 5711.29 & 5383.51 \\
         8.94637 & 2385.34 & 2395.4576 & 5704.28 & 2388.3903 & 5412.32 & 5699.14 & 5410.62 \\
         8.31297 & 2189.29 & 2193.3351 & 5708.03 & 2188.7842 & 5448.47 & 5705.90 & 5448.80 \\
         8.10751 & 2094.38 & 2103.6034 & 5698.27 & 2093.0120 & 5449.90 & 5693.36 & 5450.83 \\
         8.09994 & 2004.90 & 1998.5077 & 5648.26 & 1999.8838 & 5419.02 & 5651.76 & 5422.70 \\
         7.71463 & 2004.07 & 2009.2157 & 5713.42 & 2004.9007 & 5490.38 & 5710.72 & 5489.80 \\
	 \multicolumn{7}{c}{ML Relation from interpolated \citet{chi89}} \\
	 30.9661 & 12359.6 & 12202.760 & 5566.66 & 12354.904 & 5029.65 & 5585.98 & 5030.07 \\
	 15.9744 & 5011.60 & 5031.4017 & 5550.08 & 5024.6207 & 5173.09 & 5544.24 & 5169.84 \\
	 10.0562 & 2860.25 & 2866.2012 & 5593.28 & 2859.5002 & 5346.96 & 5590.81 & 5347.35 \\
	 8.51317 & 2447.87 & 2441.2040 & 5643.58 & 2453.1647 & 5463.35 & 5646.11 & 5460.12 \\
	 6.46303 & 1749.72 & 1749.6046 & 5719.11 & 1744.6691 & 5577.03 & 5719.24 & 5581.12 \\
         \hline 
       \end{tabular}
       }
     \end{table*} 

\begin{figure*}
       \vspace{0cm}
       \hbox{\hspace{0.2cm}\epsfxsize=7.5cm \epsfbox{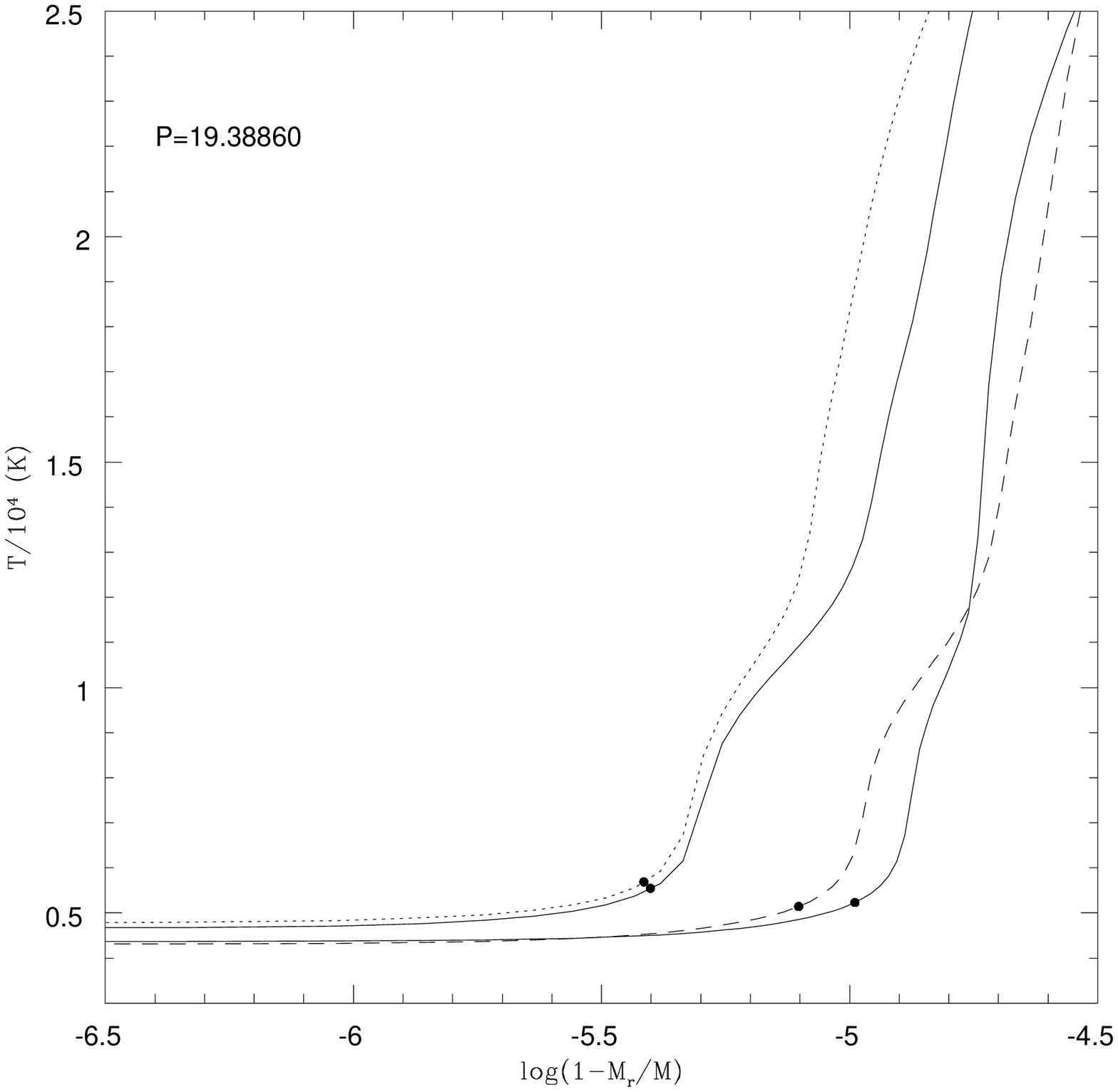}
         \epsfxsize=7.5cm \epsfbox{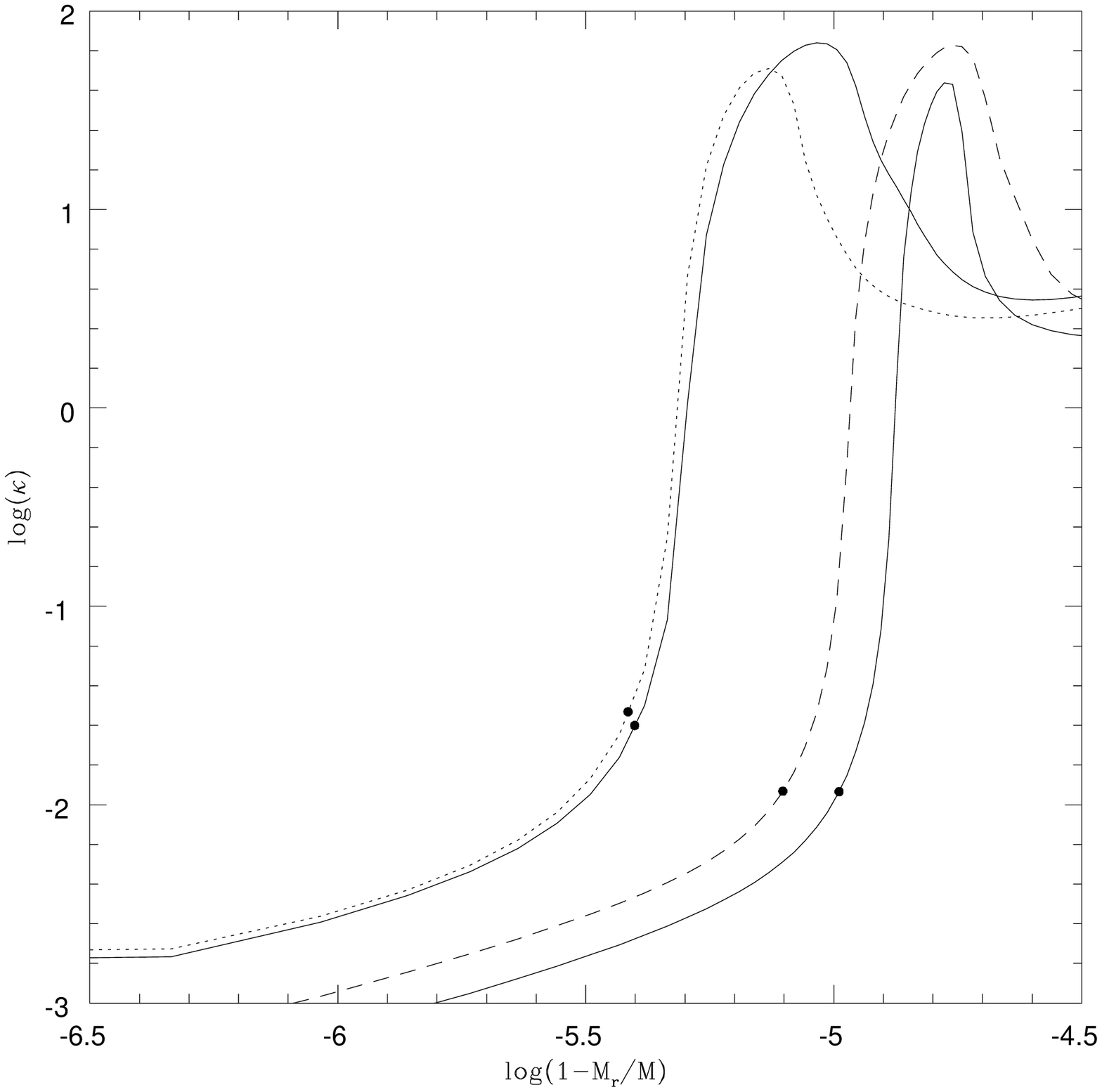}}
       \vspace{0cm}
       \caption{The temperature (left panel) and the opacity (right panel) profiles for a long period LMC model. The dotted, solid and dashed curves are for the profiles at maximum, mean and minimum light, respectively. The filled circles are the location of the photosphere at $\tau=2/3$ for each phases. The mean light profiles at the ascending and descending branch are the solid curves that lie close to the profiles at maximum light (dotted curves) and minimum light (dashed curves), respectively.}
       \label{c9bono4}
     \end{figure*}

\begin{figure*}
       \vspace{0cm}
       \hbox{\hspace{0.2cm}\epsfxsize=7.5cm \epsfbox{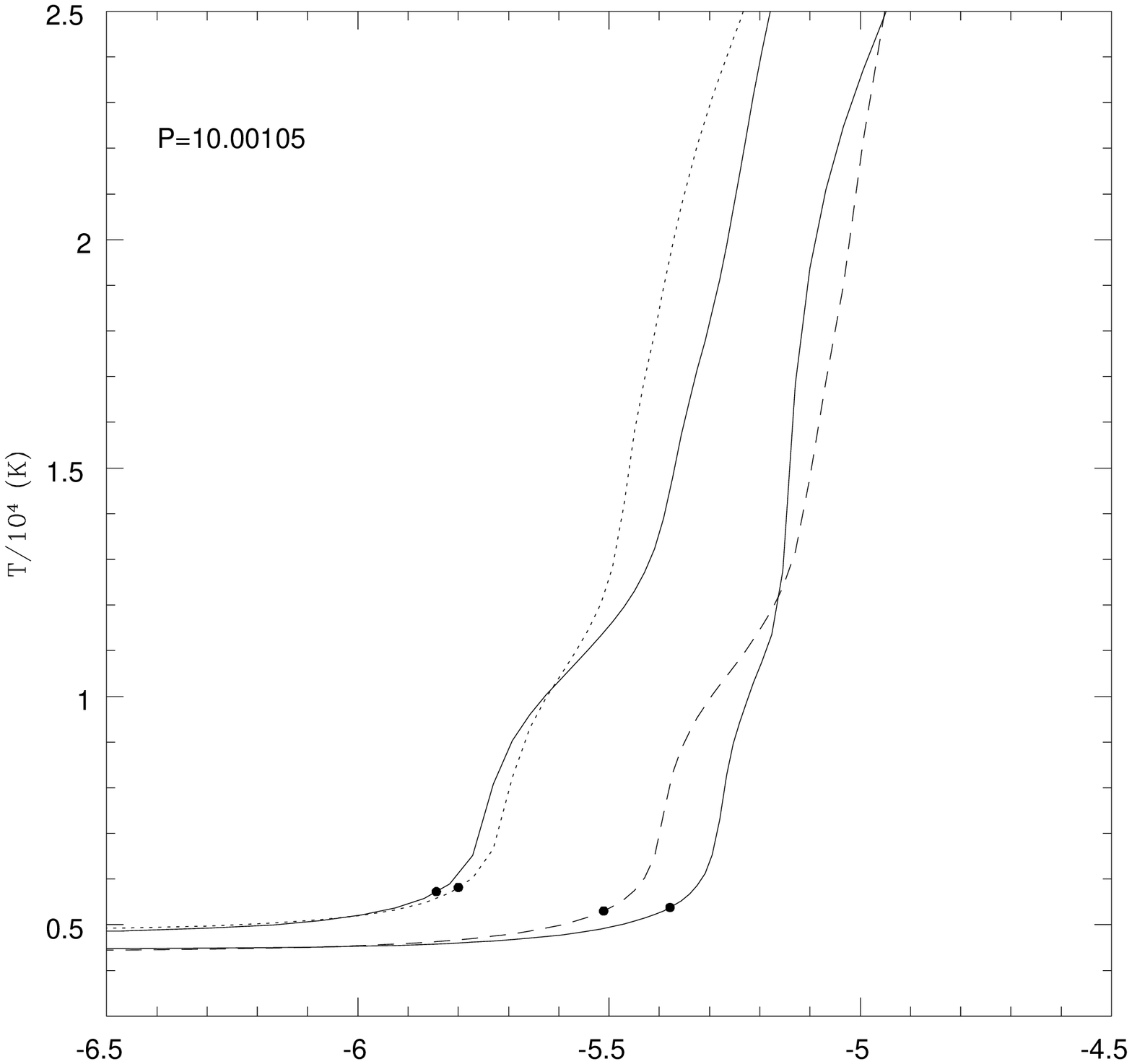}
         \epsfxsize=7.5cm \epsfbox{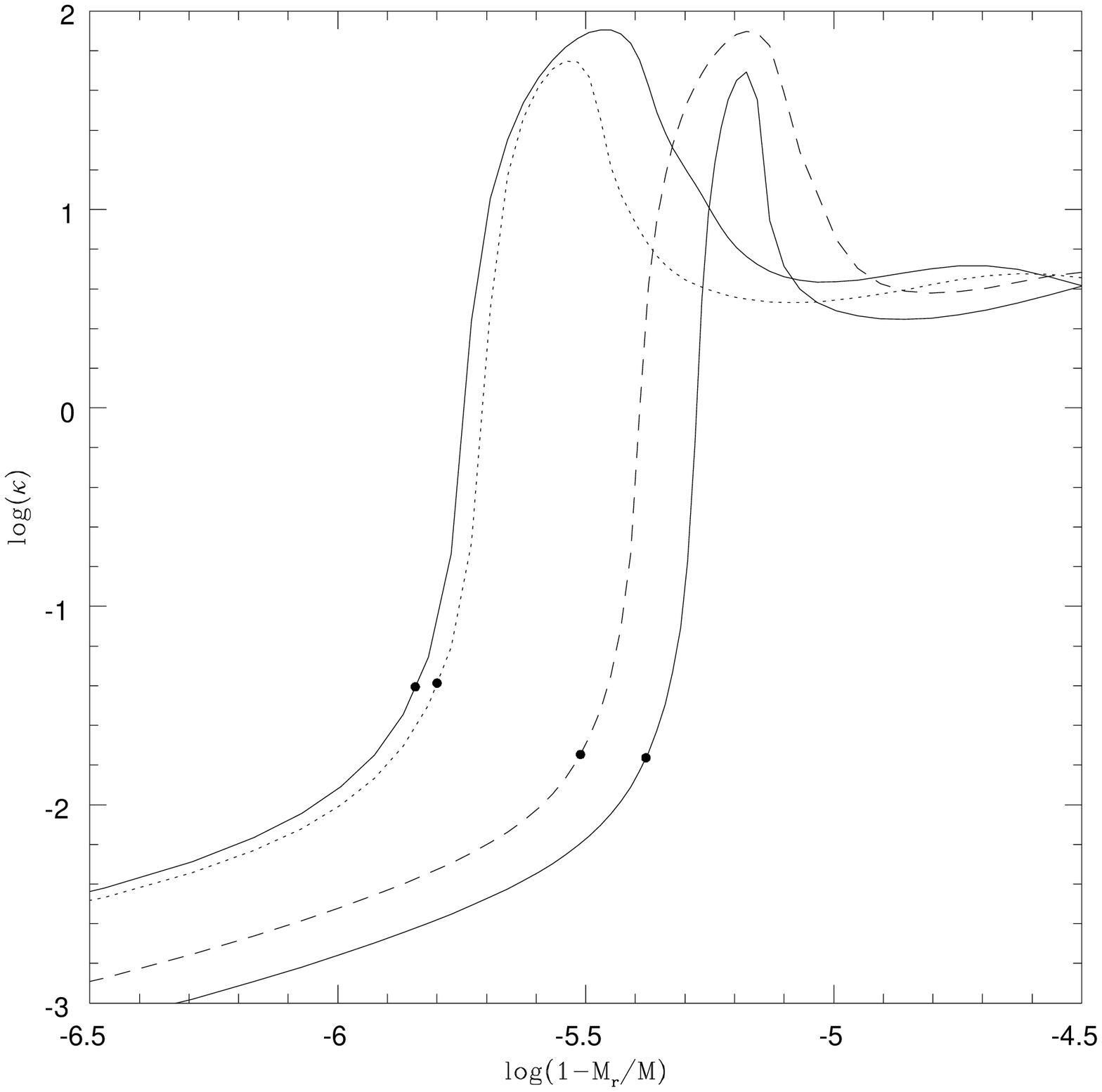}}
       \vspace{0cm}
       \caption{Same as Figure \ref{c9bono4}, but for a 10-days period LMC model.}
       \label{c9chiosi9}
     \end{figure*}

\begin{figure*}
       \vspace{0cm}
       \hbox{\hspace{0.2cm}\epsfxsize=7.5cm \epsfbox{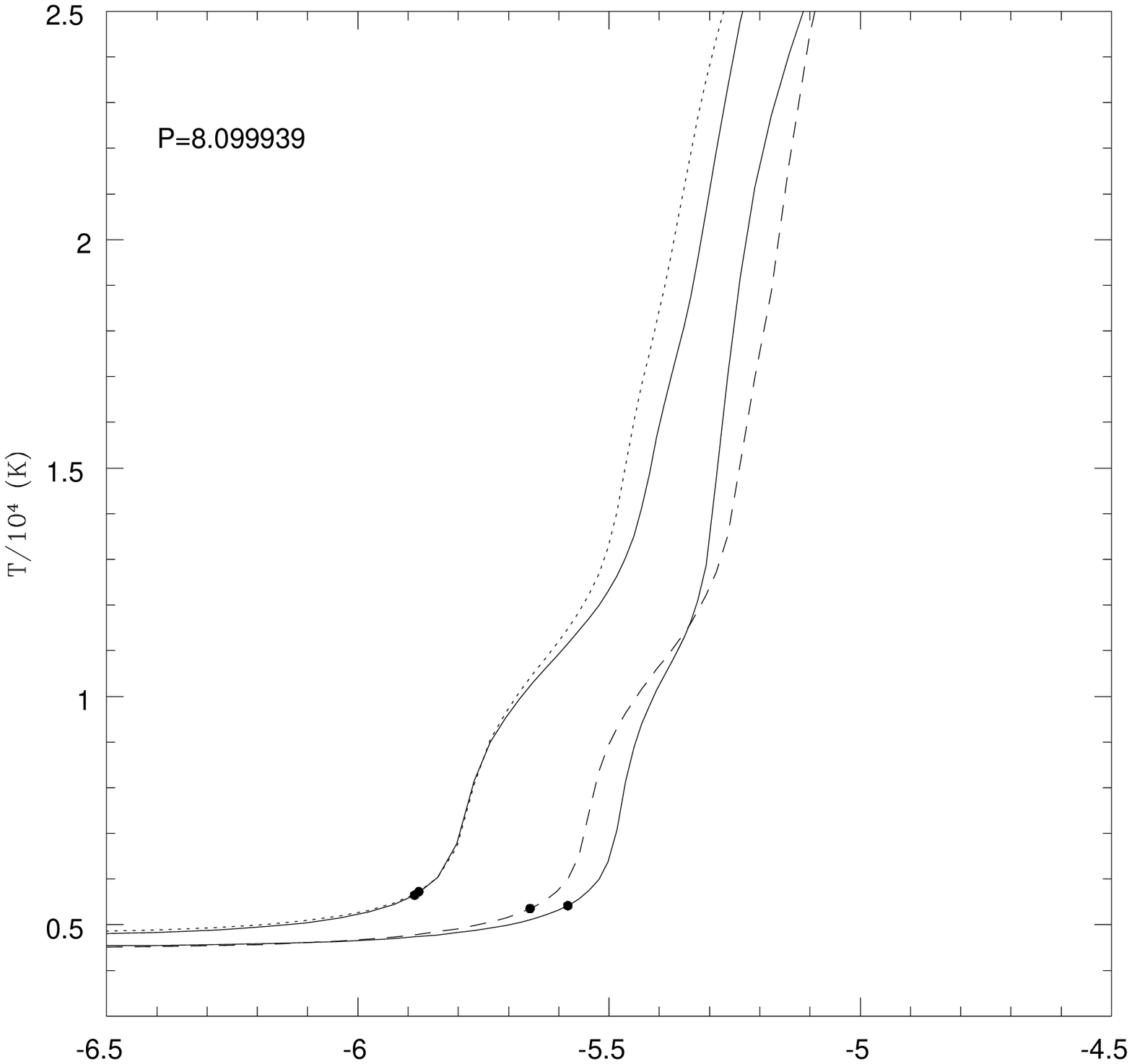}
         \epsfxsize=7.5cm \epsfbox{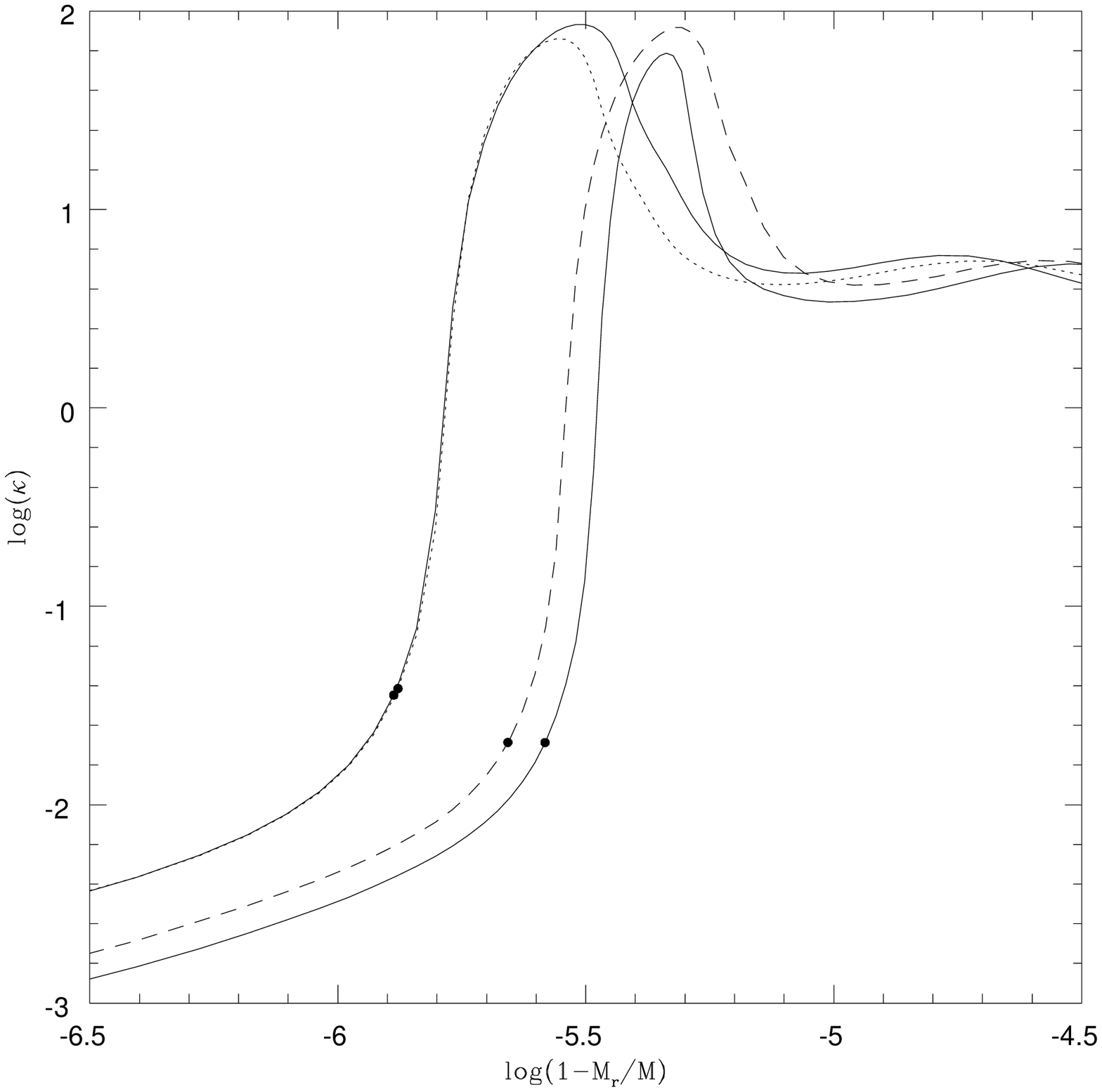}}
       \vspace{0cm}
       \caption{Same as Figure \ref{c9bono4}, but for a short period LMC model.}
       \label{c9chiosi13}
     \end{figure*}

\begin{figure*}
       \vspace{0cm}
       \hbox{\hspace{0.2cm}\epsfxsize=7.5cm \epsfbox{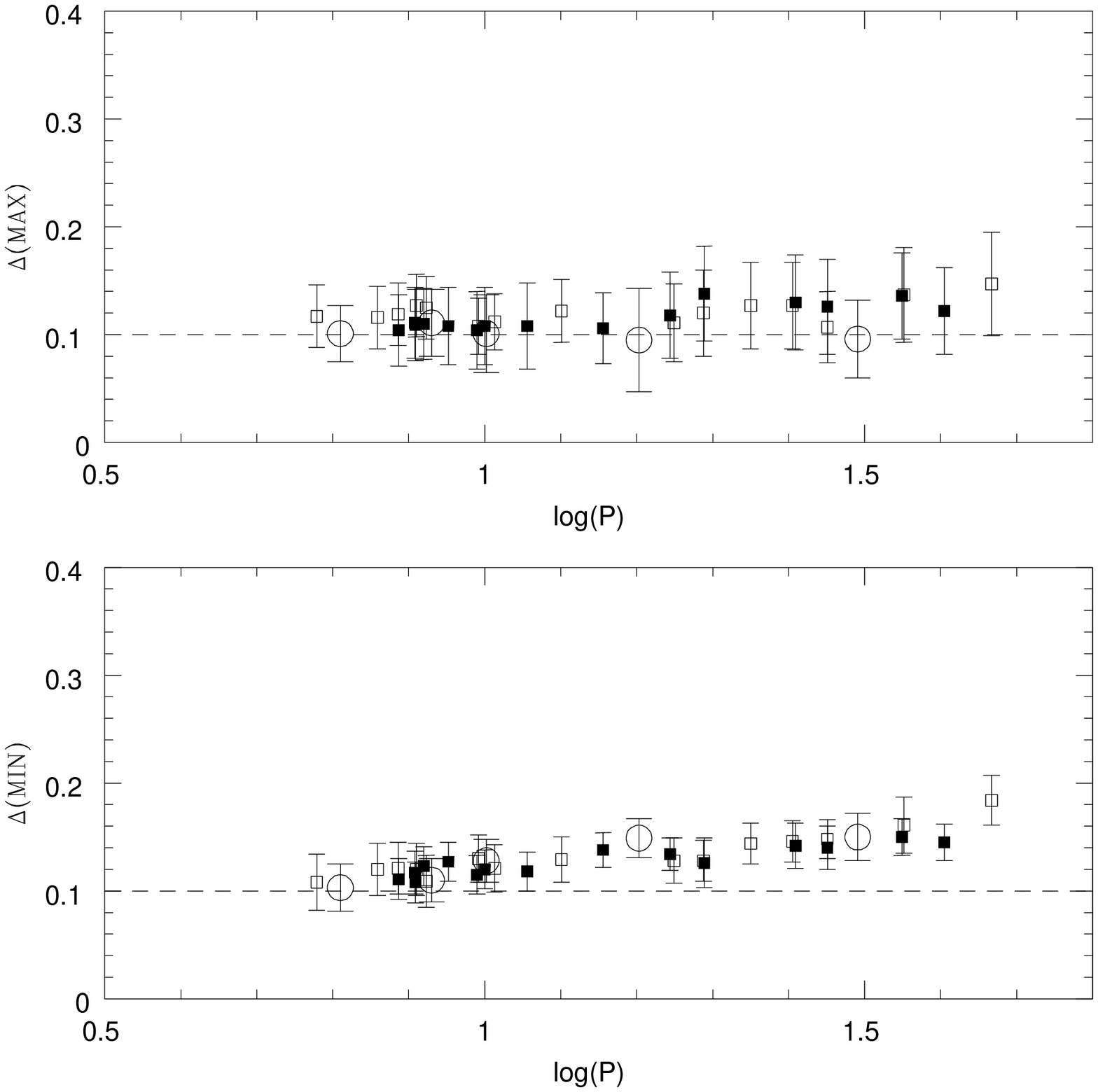}
         \epsfxsize=7.5cm \epsfbox{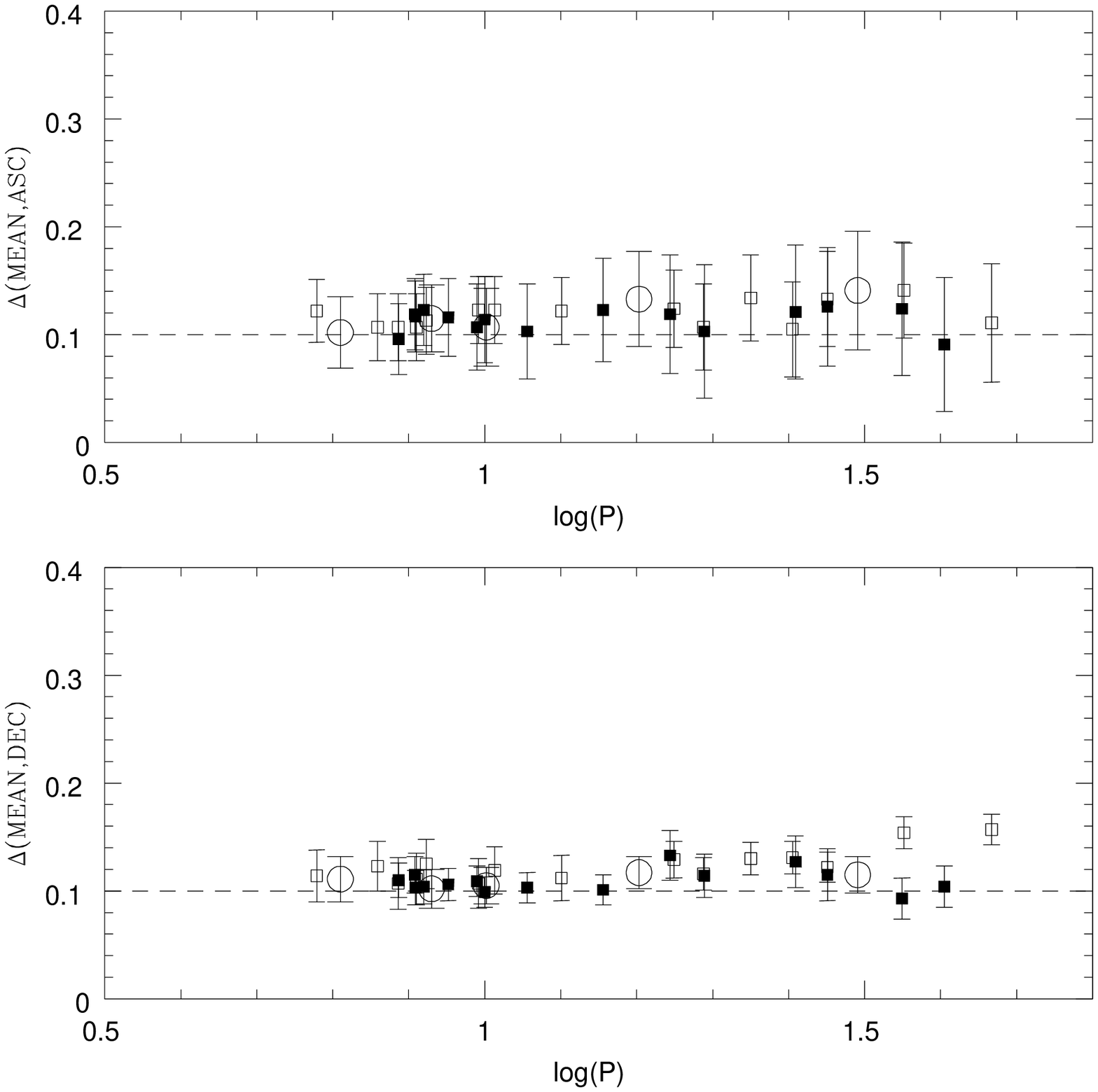}}
       \vspace{0cm}
       \caption{The plots of $\Delta$ as function of $\log(P)$. The open squares, solid squares and open circles are the models calculated with \citet{bon00} ML relation, \citet{chi89} ML relation and interpolated \citet{chi89} ML relation, respectively. The dashed lines represent (roughly) the outer boundary of the HIF.}
       \label{c9deltalmc}
     \end{figure*}

     The effective temperatures for the full amplitude models in Table \ref{tabinput} at the corresponding maximum and minimum light (or luminosity) are given in Table \ref{c9tabmaxmin}. For the effective temperatures at mean light, the temperatures for the mean light at ascending and descending branch of the light (or luminosity) curve are not the same (e.g., in Paper II), hence Table \ref{c9tabmean} gives the effective temperature at these phases for our LMC models. The layout of Table \ref{c9tabmean} is the same as table 3 from Paper II. Following Paper II,  the locations of the photosphere can be identified in the temperature and opacity profiles. These are displayed in Figure \ref{c9bono4}-\ref{c9chiosi13} with a $\log(P)>1.0$, a $\log(P)=1.0$ and a $\log(P)<1.0$ model, respectively. The left and right panels of Figure \ref{c9bono4}-\ref{c9chiosi13} are the temperature and opacity profiles respectively. The photospheres are marked as filled circles in these figures. Finally, the plots of the $\Delta$, the ``distance'' between the photosphere and the HIF from the temperature profiles, as a function of pulsating period for the LMC models are presented in Figure \ref{c9deltalmc} with the three ML relations used. In Paper II, it is found that the distribution of $\Delta$ as a function of period is almost independent of the adopted ML relation. This is also seen in the LMC models as depicted in Figure \ref{c9deltalmc}.

%**********************************************************
%      FIGURE: long period model
%**********************************************************
 
%**********************************************************
%      FIGURE: 10d model
%**********************************************************
 
%**********************************************************
%      FIGURE: short period model
%**********************************************************
 
%**********************************************************
%      FIGURE: LMC delta
%**********************************************************
 
     Figure \ref{c9bono4}-\ref{c9chiosi13} and Figure \ref{c9deltalmc} bear witness to the fact that at maximum light, the photosphere lies at the base of the HIF for all of the models. Although there is a slight deviation for some longer period models, the location of the photosphere is close to the HIF within the error bars (which are defined as the coarseness of the grid points around the location of the HIF). As in Paper II for the Galactic models, the closeness of the photosphere to the base of the HIF, for reasonably low densities, results in a flat or almost flat PC relation for the long period LMC Cepheids. In the case of minimum light, even though Figure \ref{c9deltalmc} implies that $\Delta (MIN)$ is nearly constant across the period range and the photosphere is near the base of the HIF, as in the case of maximum light, $\Delta (MIN)$ does follow a shallow correlation with period after 10 days. Judging from the error bars of $\Delta (MIN)$ and from Figure \ref{c9bono4}-\ref{c9chiosi13}, there is tentative evidence that the photosphere is disengaged from the HIF for $\log(P)>1.0$ at minimum light. Hence the temperatures or the colours at minimum light are more dependent on period for $\log(P) > 1.0$ \footnote{The slopes of the $\Delta$-$\log(P)$ relation may not be correlated with the slopes of the PC relation} and the global properties.

\begin{figure*}
       \vspace{0cm}
       \hbox{\hspace{0.2cm}\epsfxsize=7.5cm \epsfbox{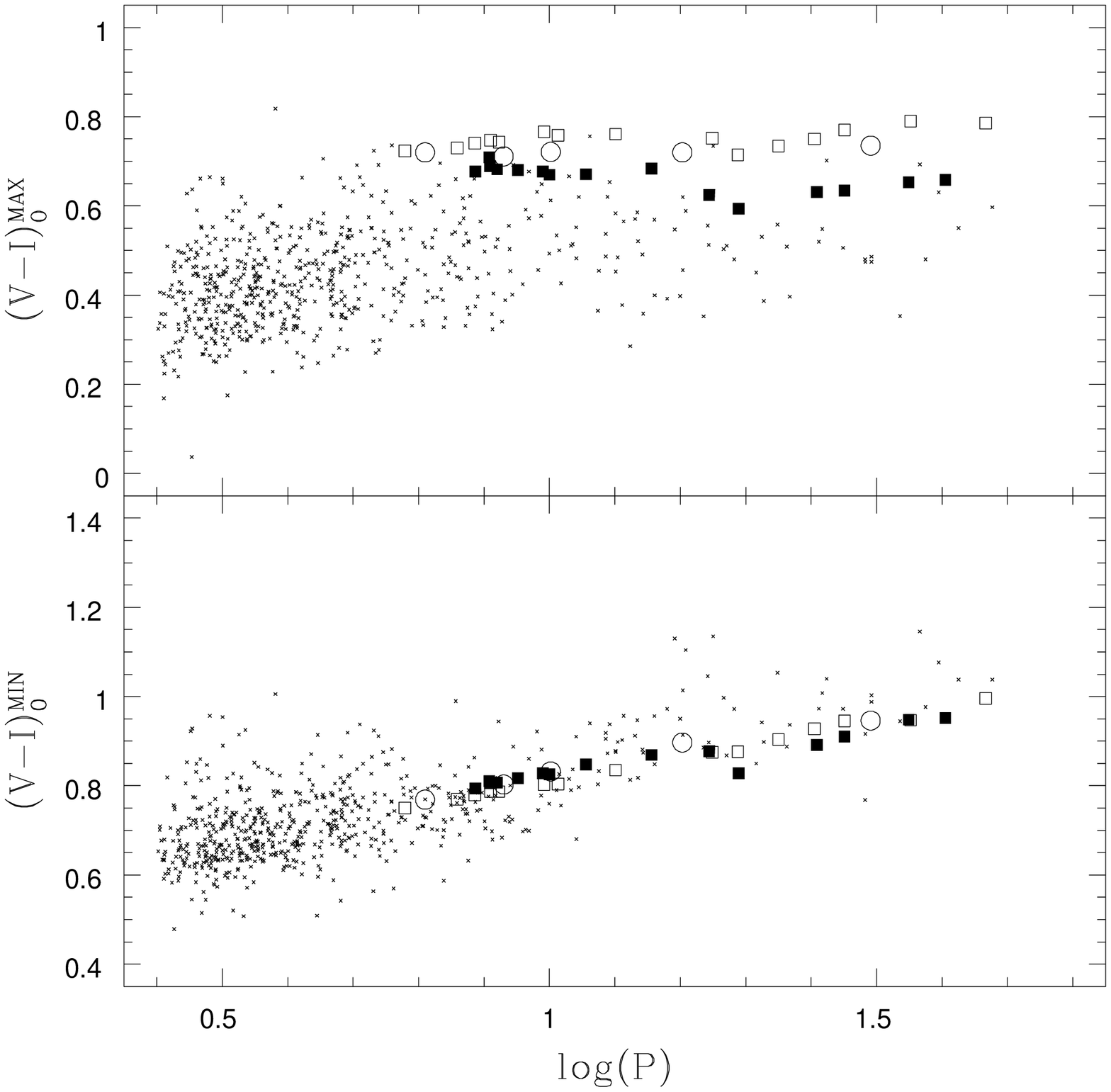}
         \epsfxsize=7.5cm \epsfbox{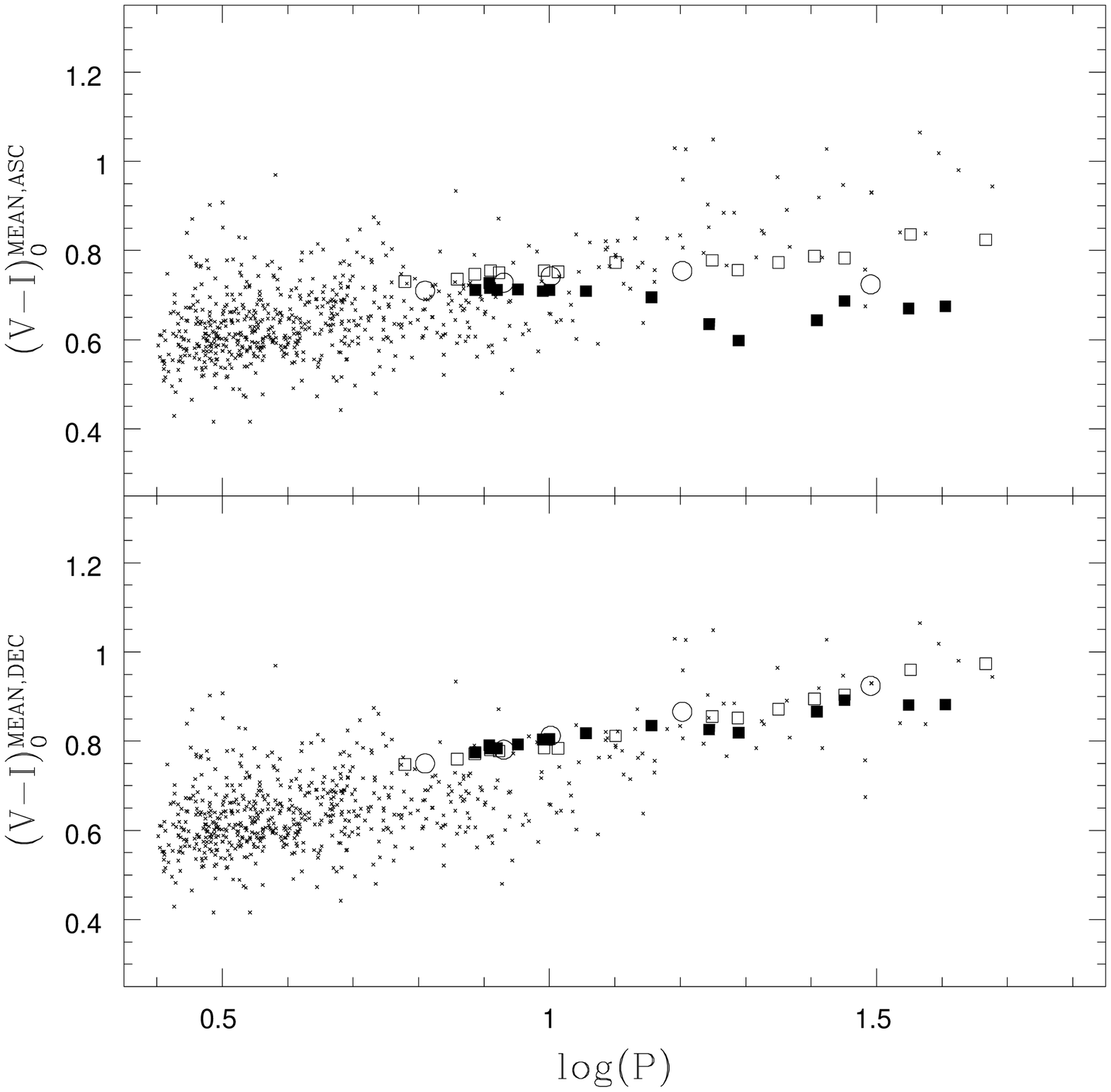}}
       \vspace{0cm}
       \caption{The LMC PC relations with the results from the models. The small crosses are the observed data points as given in Figure \ref{c9figpc}. The open squares, solid squares and open circles are the models calculated with \citet{bon00}, \citet{chi89} and interpolated \citet{chi89} ML relations, respectively. The temperatures of the models are converted to the $(V-I)$ colour using the {\tt BaSeL} database. Left panel: PC relations at maximum and minimum light. Right panel: PC relations at mean light for both of the ascending and descending means. The mean colours for the observed data are the $(V-I)_0^{\mathrm{phmean}}$ as given in Figure \ref{c9figpc}.}
       \label{c9modelpc}
     \end{figure*}

\begin{figure*}
       \vspace{0cm}
       \hbox{\hspace{0.2cm}\epsfxsize=7.5cm \epsfbox{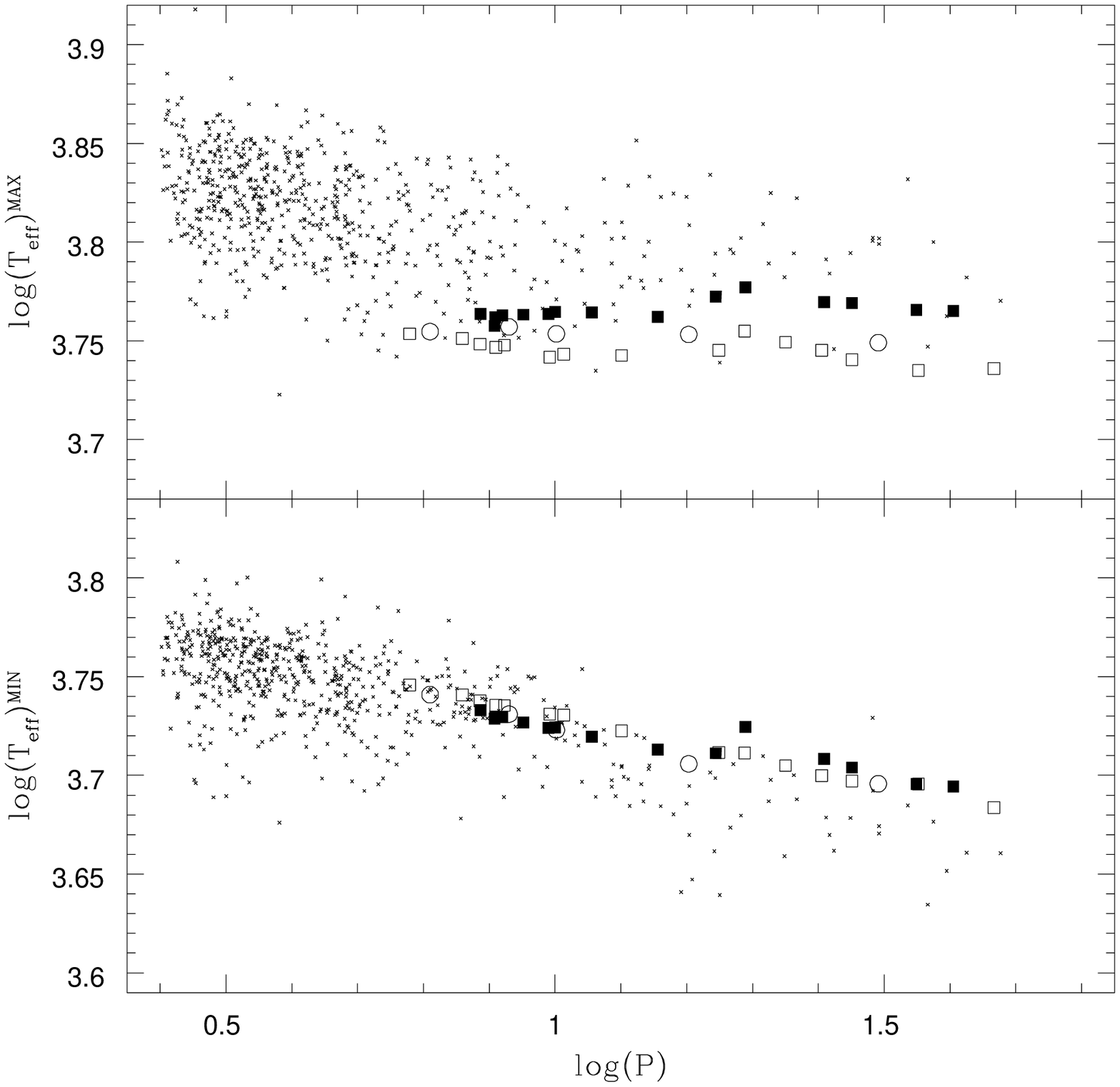}
         \epsfxsize=7.5cm \epsfbox{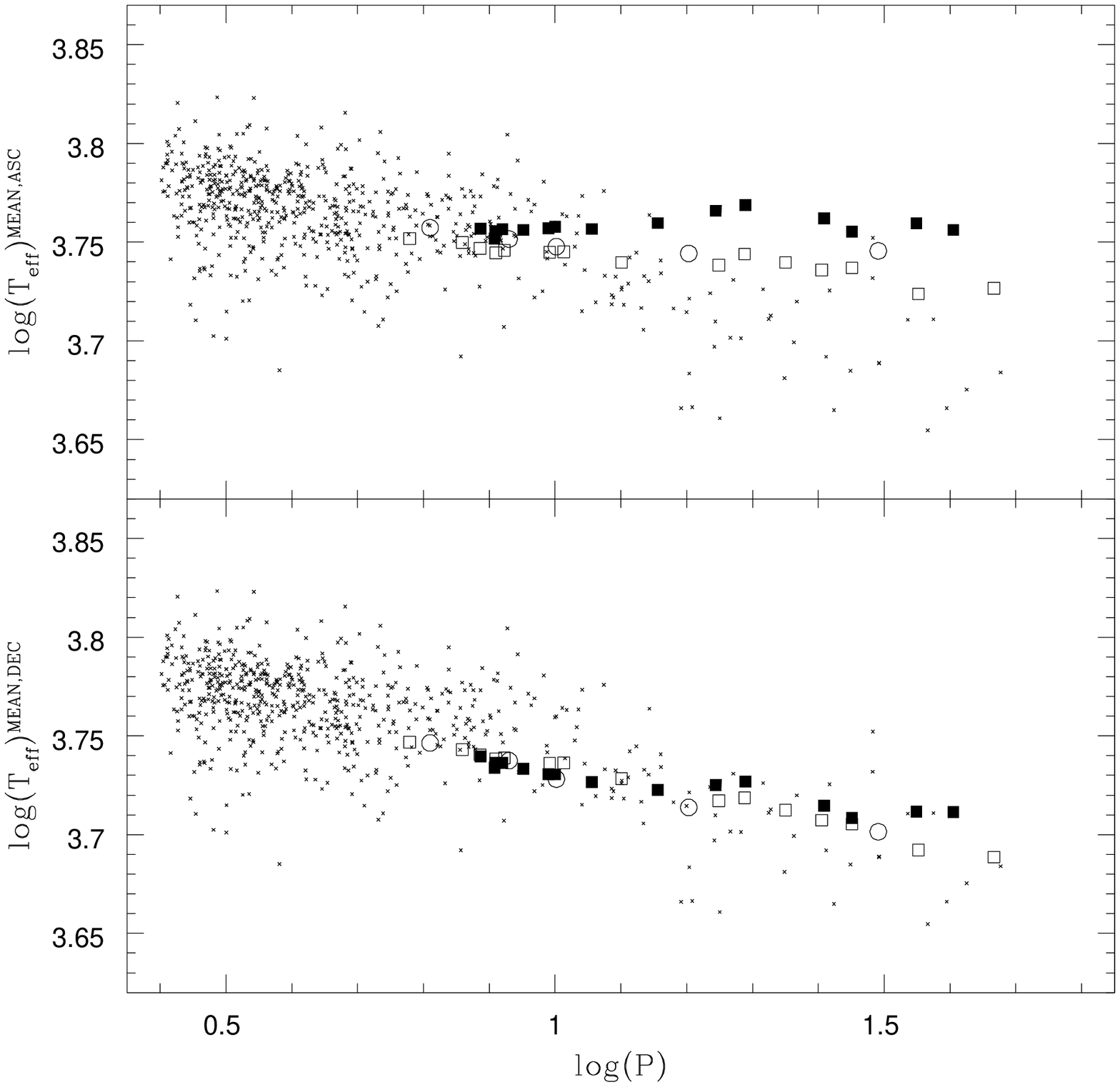}}
       \vspace{0cm}
       \caption{The plots of $\log(T)$-$\log(P)$ relations for the LMC data and models. The symbols are the same as in Figure \ref{c9modelpc}. The conversion of the $(V-I)$ colours to the temperature are done using the equations given in \citet{bea01}. Left panel: $\log(T)$-$\log(P)$ relations at maximum and minimum light. Right panel: $\log(T)$-$\log(P)$ relations at mean light for both of the ascending and descending means. }
       \label{c9modelpt}
     \end{figure*}

\begin{figure*}
       \vspace{0cm}
       \hbox{\hspace{0.2cm}\epsfxsize=7.5cm \epsfbox{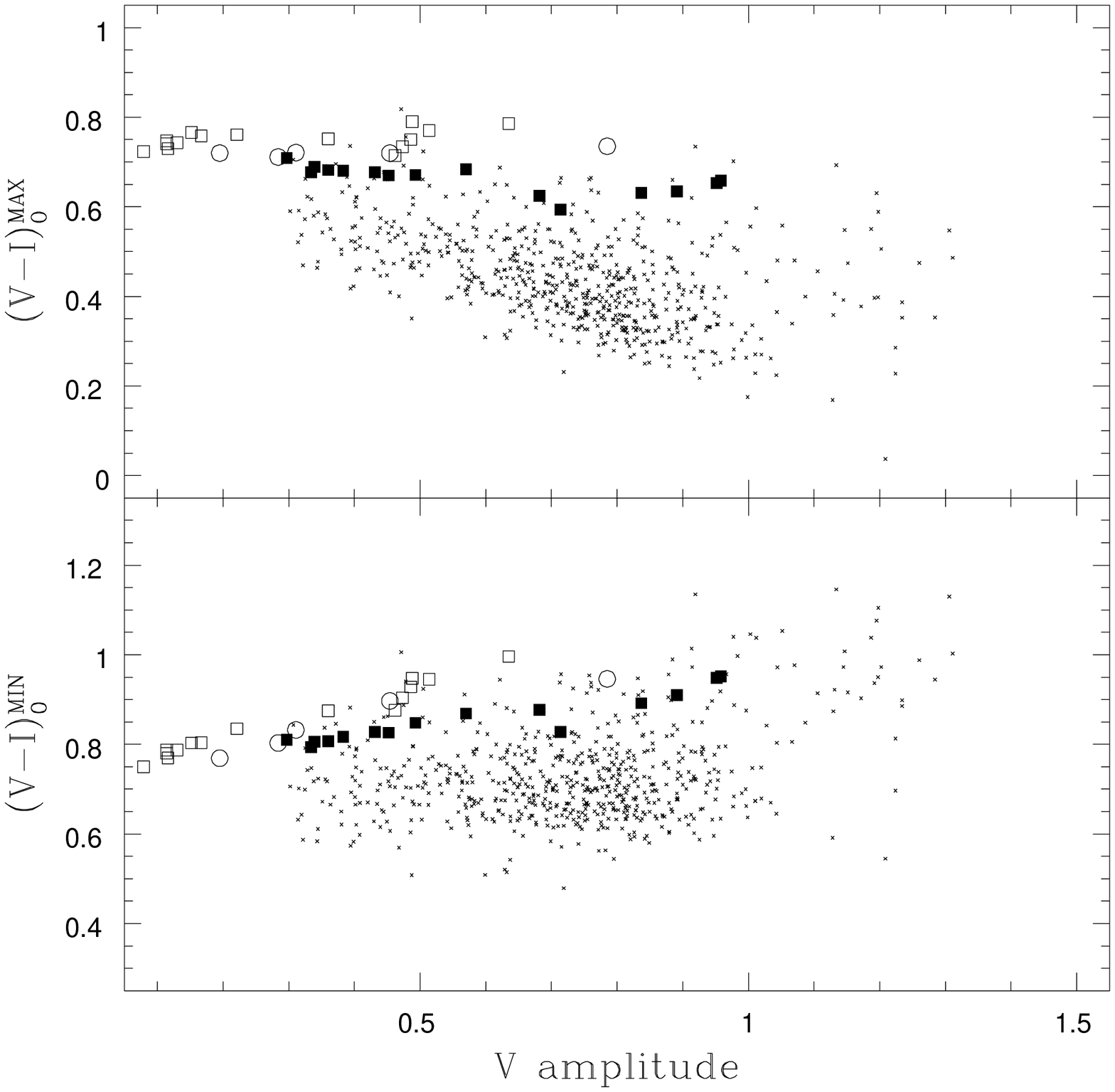}
         \epsfxsize=7.5cm \epsfbox{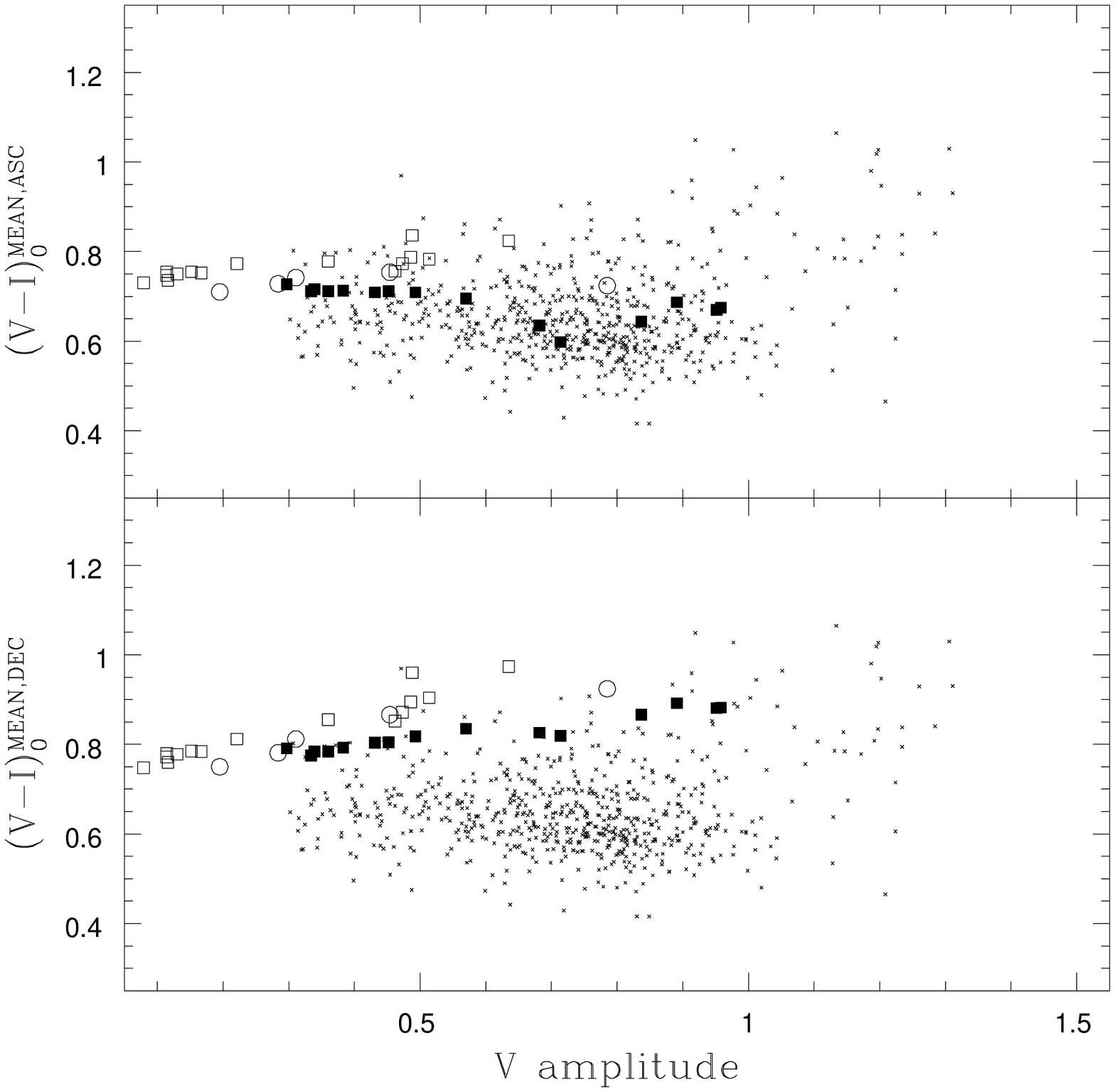}}
       \vspace{0cm}
       \caption{The LMC AC relations with the results from the models. The small crosses are the observed data points as given in Figure \ref{c9figac}. The open squares, solid squares and open circles are the models calculated with \citet{bon00}, \citet{chi89} and interpolated \citet{chi89} ML relations, respectively. The bolometric light curves from models are converted to $V$-band light curves with the $BC$ obtained from the {\tt BaSeL} database. Left panel: AC relations at maximum and minimum light. Right panel: AC relations at mean light for both of the ascending and descending means. The mean colours for the observed data are the $(V-I)_0^{\mathrm{phmean}}$ as given in Figure \ref{c9figac}.}
       \label{c9modelac}
     \end{figure*}

\begin{figure*}
       \vspace{0cm}
       \hbox{\hspace{0.2cm}\epsfxsize=7.5cm \epsfbox{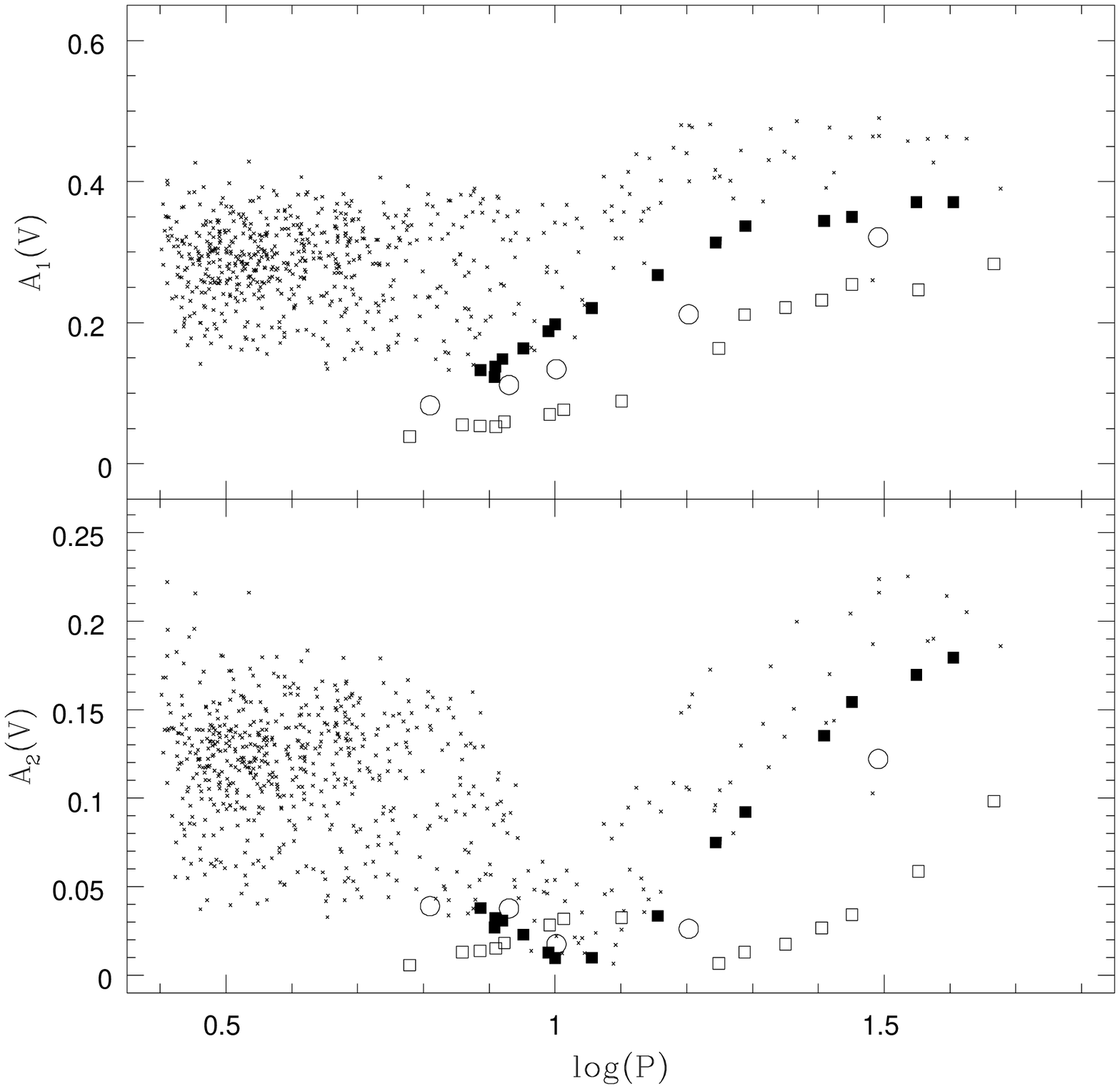}
         \epsfxsize=7.5cm \epsfbox{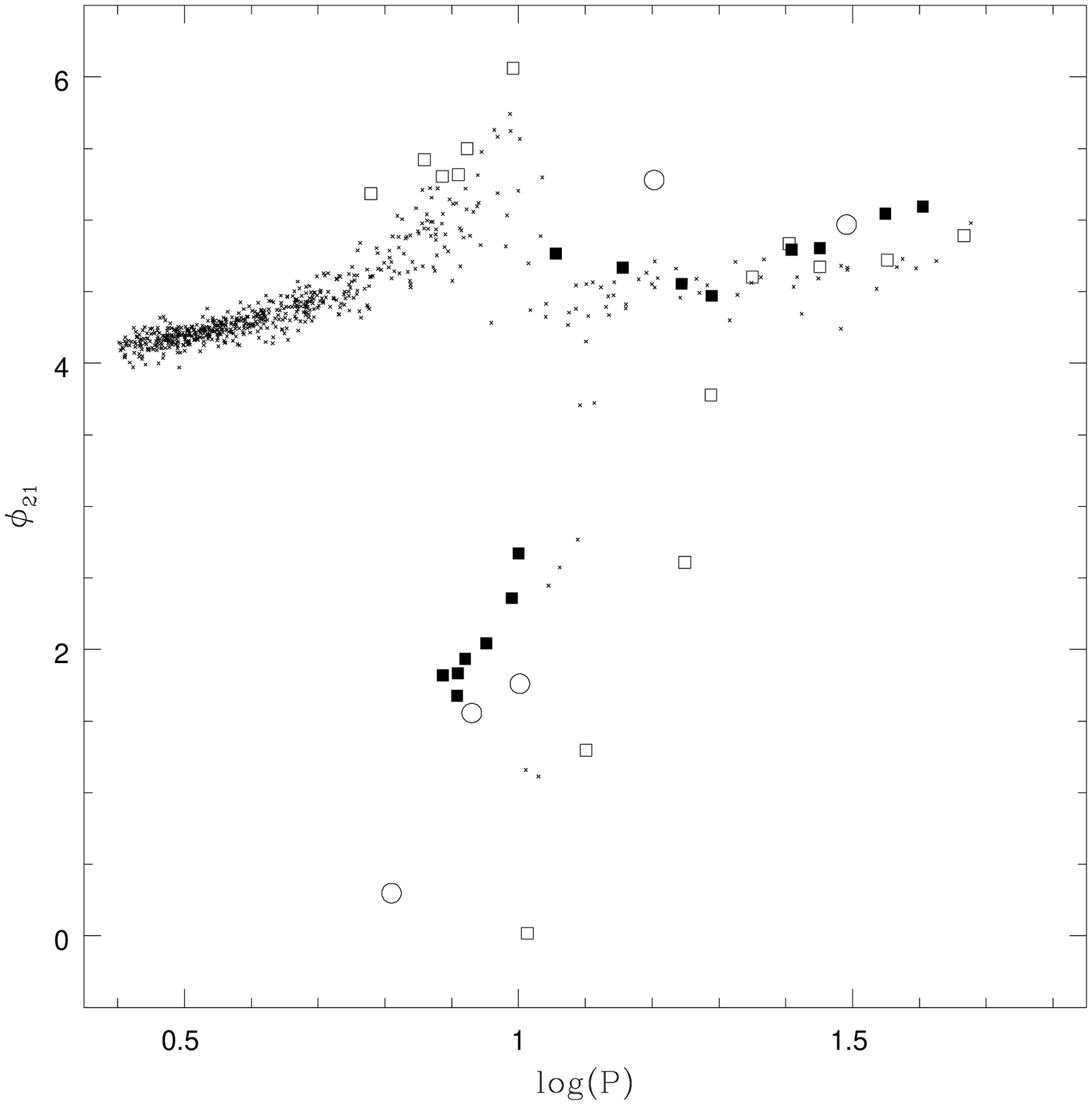}}
       \vspace{0cm}
       \caption{The Fourier parameters as function of periods for the LMC observational data (small crosses) and the models, where open squares, solid squares and open circles are the models calculated with \citet{bon00}, \citet{chi89} and interpolated \citet{chi89} ML relations, respectively. Upper Left: Plot of $A_1$ vs $\log(P)$ in $V$-band. Lower Left: Plot of $A_2$ vs $\log(P)$ in $V$-band. Right: Plot of $\phi_{21}$ vs $\log(P)$, where $\phi_{21}=\phi_2-2\phi_1$ \citep{sim81,nge03}.}
       \label{c9lmcfourier}
     \end{figure*}

    Theoretical quantities that can be computed from the models and compared with data include the pulsation periods, the $V$-band amplitudes and the Fourier parameters, the temperatures and colours at the maximum, mean and minimum light. These are the PC plots, the AC plots, the period-temperature plots and the Fourier parameters plots portrayed in Figures \ref{c9modelpc}-\ref{c9lmcfourier}. The temperatures in Table \ref{c9tabmaxmin} \& \ref{c9tabmean}, after conversion to the $(V-I)$ colours as mentioned in previous section, are superimposed along with the observed LMC PC relations as plotted in Figure \ref{c9modelpc}. Similarly, Figure \ref{c9modelpt} graphs the same quantities but on the $\log(T)$-$\log(P)$ plane with the observed $(V-I)$ colours converted to temperatures using the prescriptions given in Section 5. The theoretical bolometric light curves are converted to the $V$-band light curves with the bolometric corrections obtained from the {\tt BaSeL} database mentioned previously. From the theoretical $V$-band light curves, the amplitudes can be estimated and these are displayed in Figure \ref{c9modelac} along with the colours from models to compare with the empirical AC relations. The Fourier parameters of the theoretical $V$-band light curves can also be obtained with ($n=6$) Fourier expansion. These Fourier parameters are compared with the observational data in Figure \ref{c9lmcfourier}. Several features are noticed from Figure \ref{c9modelpc}-\ref{c9lmcfourier}: 

    \begin{enumerate}
    \item The general trends of the models qualitatively match the observational data. There are greater discrepancies between the data and short period models, particularly in matching the observed light curve amplitudes.
    \item The models with the ML relation from \citet{chi89}, with lower $L/M$ ratio, do better in matching the observations. These models also tend to lie near the envelopes of the PC, AC, $\log(T)$-$\log(P)$ and $A_n(V)$-$\log(P)$ relations defined by the observational data. 
    \item The slopes of the period-colour (or period-temperature) relations at maximum and minimum light from the models roughly match the observational data, i.e., the theoretical PC(max) relation is approximately flat and there is a relation at minimum light.   
    \item The temperatures from the models with the \citet{bon00} ML relation is cooler (hence redder) than the models with the \citet{chi89} ML relation and the observed data at maximum light. In contrast, the temperatures (or the colours) at minimum light from the models with these two ML relations are consistent with each other and are located near the blue edge of the observed data. 
    \item The means at the descending branches are in better agreement with the observed data than the means at the ascending branches. This is because the observed means, $(V-I)_0^{\mathrm{phmean}}$, are obtained mostly from the descending branches. Though previous researchers have noted that temperatures on the ascending and descending branches are not the same at mean light (as Cepheids exhibit loops in CMD), what is new here is the way the nature of the HIF changes during the pulsation.  
    \item The behaviors of the models from the interpolated \citet{chi89} ML are closer to the models from \citet{bon00} ML relation because their slopes are very similar.
    \item The amplitudes of the theoretical light curves (in both of the bolometric and $V$-band light curves) are smaller than the observations at given period, especially for the models with the \citet{bon00} ML relation. These can be seen from the AC relations as given in Figure \ref{c9modelac} and the left panels of Figure \ref{c9lmcfourier}.
    \end{enumerate}

    \ni Overall, some agreements and disagreements are found between the theoretical quantities and the observational data. It is also found out that there are some problems associated with the pulsation codes when the LMC models are constructed: these include the smaller amplitude of the model light curves and the cooler temperatures at the maximum light (especially with \citealt{bon00} ML relation). Note that from equation (1), cooler temperatures at maximum light imply that the amplitudes will be lower at given period. Varying other parameters in the pulsation codes, including the $\alpha$ parameters, does not improve the situation, though perhaps a more detailed and systematic study of the dependence of LMC Cepheid pulsation models on the ${\alpha}$ parameters could resolve this situation. However, we believe that the qualitative nature of the photosphere-HIF interactions as given in Figure \ref{c9deltalmc} will still hold even in models which fare better in mimicking observed amplitudes. This is in part because Figure \ref{c9deltalmc} suggest that the behaviors of $\Delta$ as a function of period are nearly, though not completely, independent of amplitudes, as the models with \citet{chi89} ML relation have higher amplitudes (although still smaller than the observations) than the models with the \citet{bon00} ML relation. However, better codes that fix these problems or the 3-D convection codes are needed in the future studies.

%**********************************************************
%      FIGURE: PC relation with models
%**********************************************************
 
%**********************************************************
%      FIGURE: PT relation with models
%**********************************************************
 
%**********************************************************
%      FIGURE: AC relation with models
%**********************************************************
 
%**********************************************************
%      FIGURE: Fourier relation with models
%**********************************************************
 
\subsection{Comparison with the Galactic Models}

%**********************************************************
%      FIGURE: compare temperature profiles
%**********************************************************
 
     \begin{figure*}
       \vspace{0cm}
       \hbox{\hspace{0.2cm}\epsfxsize=7.5cm \epsfbox{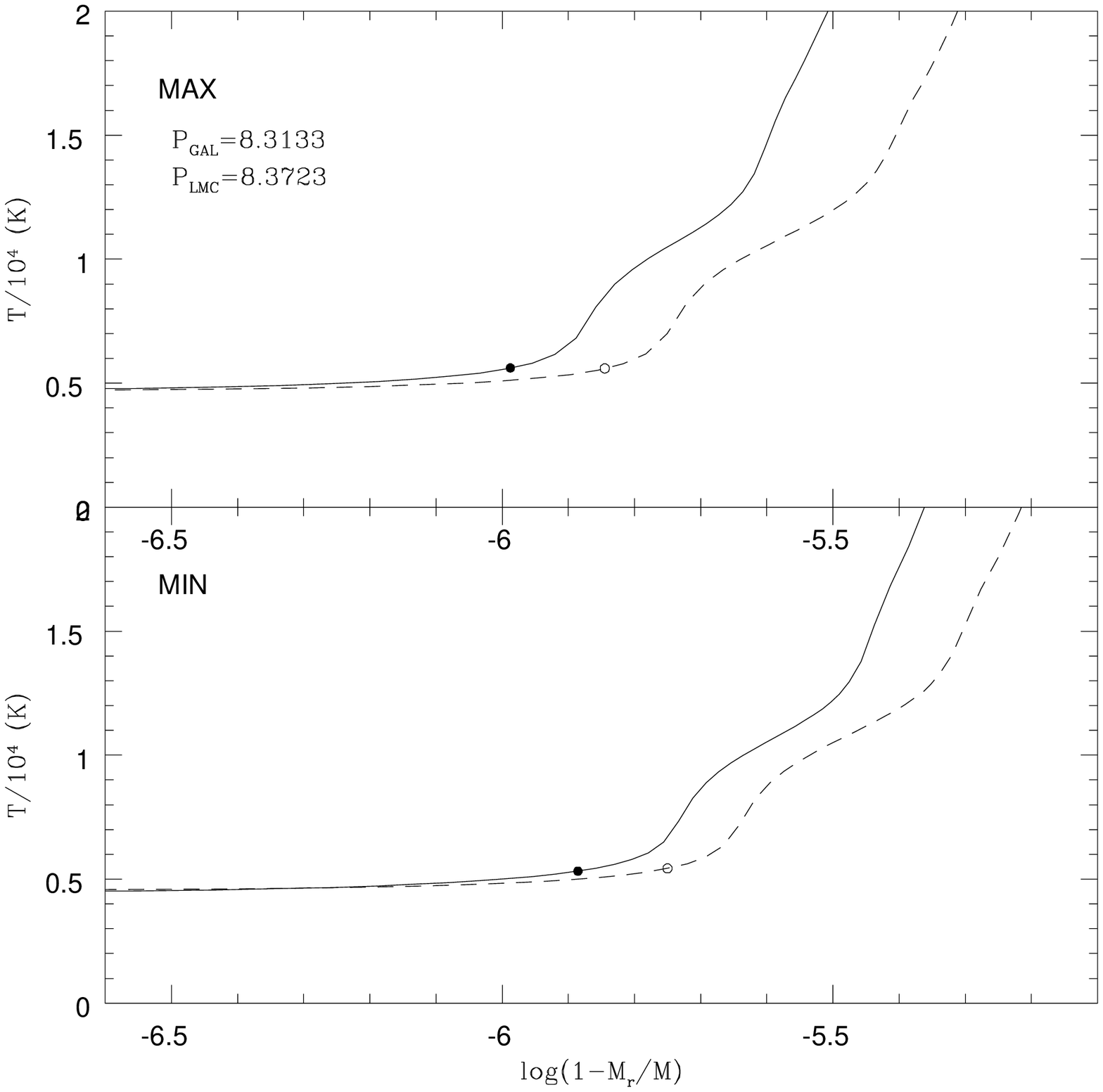}
         \epsfxsize=7.5cm \epsfbox{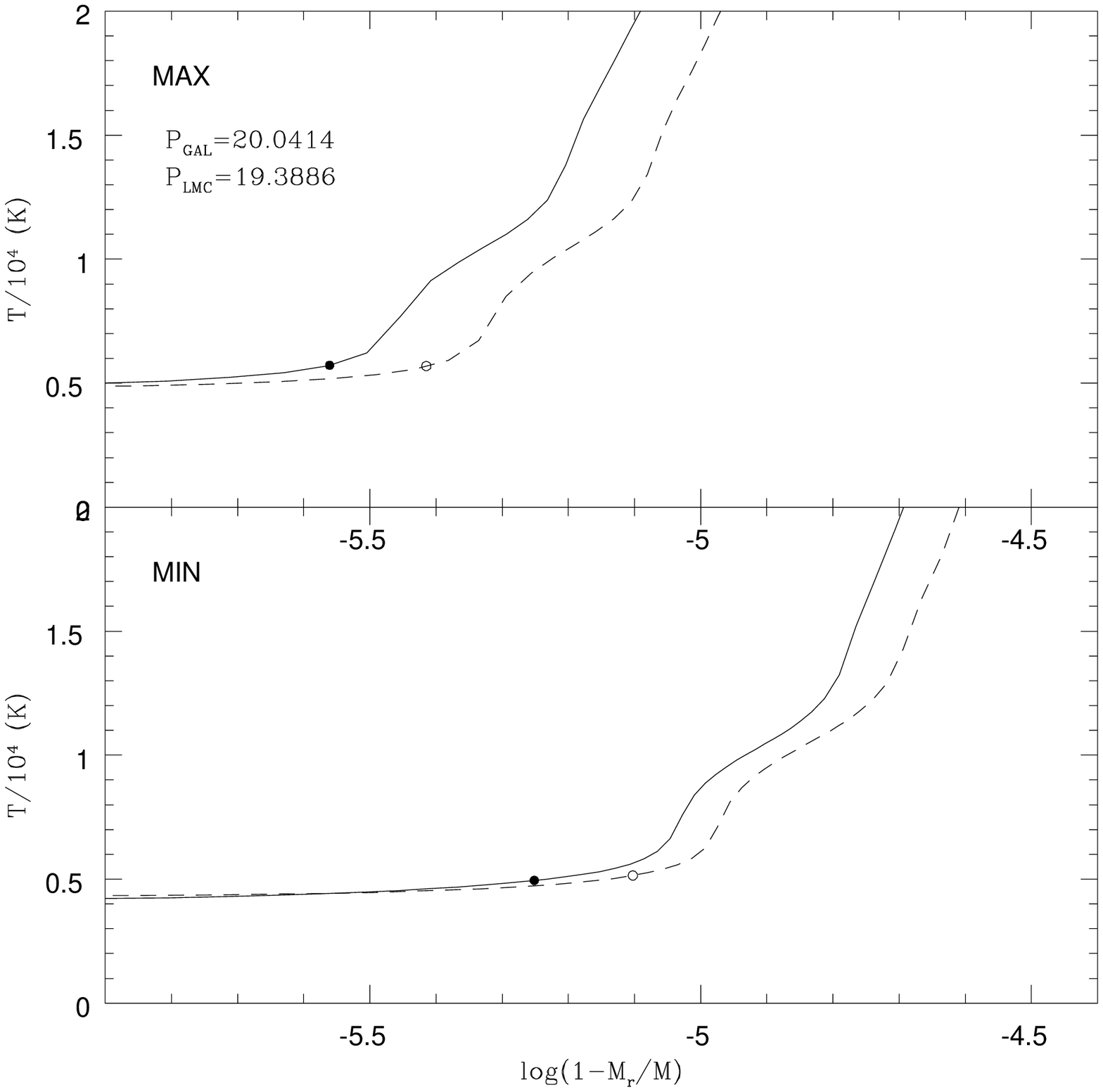}}
       \vspace{0cm}
       \caption{Comparisons of the temperature profiles between the Galactic and LMC models at maximum (upper panel) and minimum (lower panel) light. The solid and dashed curves represent the temperature profiles from the Galactic and the LMC models, respectively. The left and right panels are for the short and long period models, respectively, with the periods labeled at the upper left corners. The filled (for the Galactic models) and open (for the LMC models) circles are the locations of the photosphere in these models.}
       \label{c9pt}
     \end{figure*}

%**********************************************************
%      FIGURE: delta compare
%**********************************************************
 
     \begin{figure*}
       \vspace{0cm}
       \hbox{\hspace{0.2cm}\epsfxsize=7.5cm \epsfbox{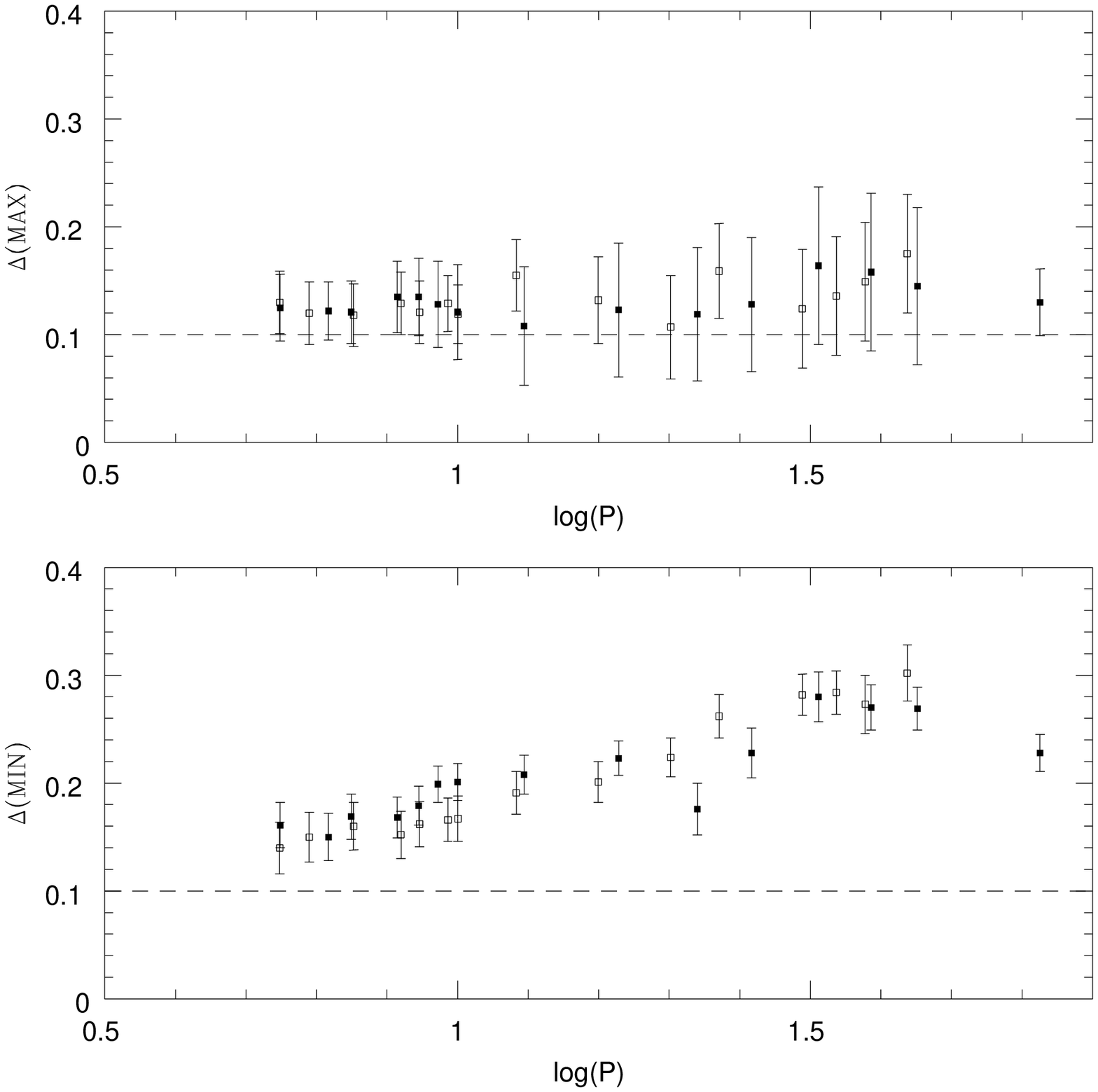}
         \epsfxsize=7.5cm \epsfbox{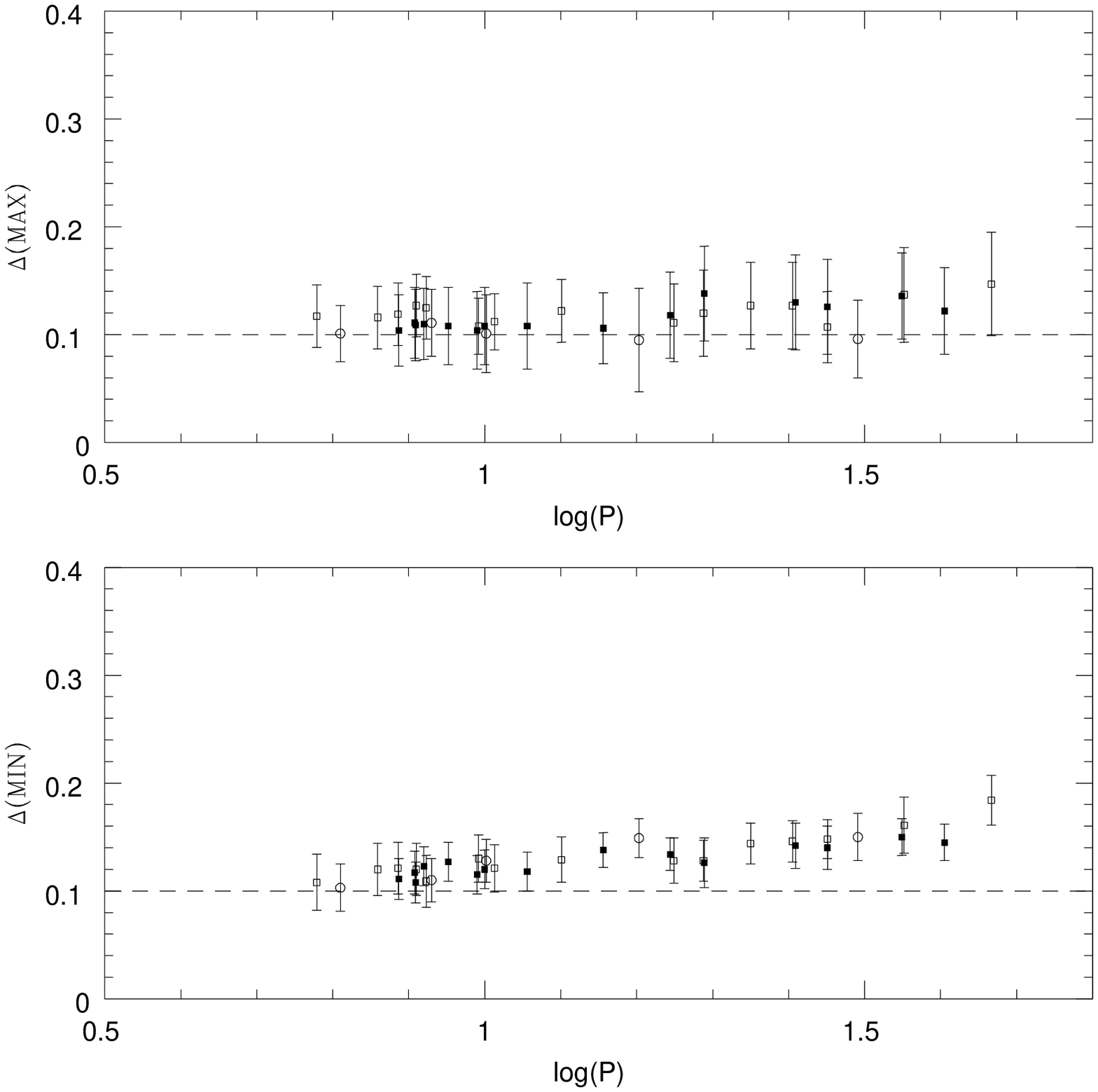}}
       \vspace{0cm}
       \caption{Comparisons of the $\Delta$-$\log(P)$ plots between the Galactic (left panel) and the LMC (right panel) models at maximum (top panel) and minimum (bottom panel) light.}
       \label{c9delta}
     \end{figure*}

     The temperature profiles from the Galactic models given in Paper II and the LMC models are compared in Figure \ref{c9pt} at maximum and minimum light. The upper panels of Figure \ref{c9pt} suggest that at maximum light, the photosphere is not far from the base of the HIF in both of the Galactic and the LMC models. In contrast, the photosphere is further away from the HIF in the Galactic models than the LMC models at minimum light. The HIF is located further out in the mass distribution for the Galactic models. The plots of the $\Delta$-$\log(P)$ relation from the Galactic and LMC models at maximum and minimum light are also compared in Figure \ref{c9delta}. It can be seen from the figure that at maximum light, the behavior of both Galactic and LMC models is similar, where the photosphere is near the base of the HIF. At minimum light, the long period models show that the photosphere is disengaged from the HIF, while the behavior of the short period models is different between the Galactic and LMC models. The photosphere of the short period LMC models seems to be located closer to the HIF at minimum light, but it is not the case for the short period Galactic models. This could lead to shallower slopes of the PC(min) relation seen in the LMC Cepheids as compared to the Galactic counterparts. In terms of the HIF-photosphere interaction, there is some tentative evidence from the models that the LMC long period Cepheids behave like the Galactic Cepheids, while the short period LMC Cepheids behave like the RR Lyrae stars at minimum light.
     
%**********************************************************
%      FIGURE: density
%**********************************************************
 
     \begin{figure*}
       \vspace{0cm}
       \hbox{\hspace{0.2cm}\epsfxsize=7.5cm \epsfbox{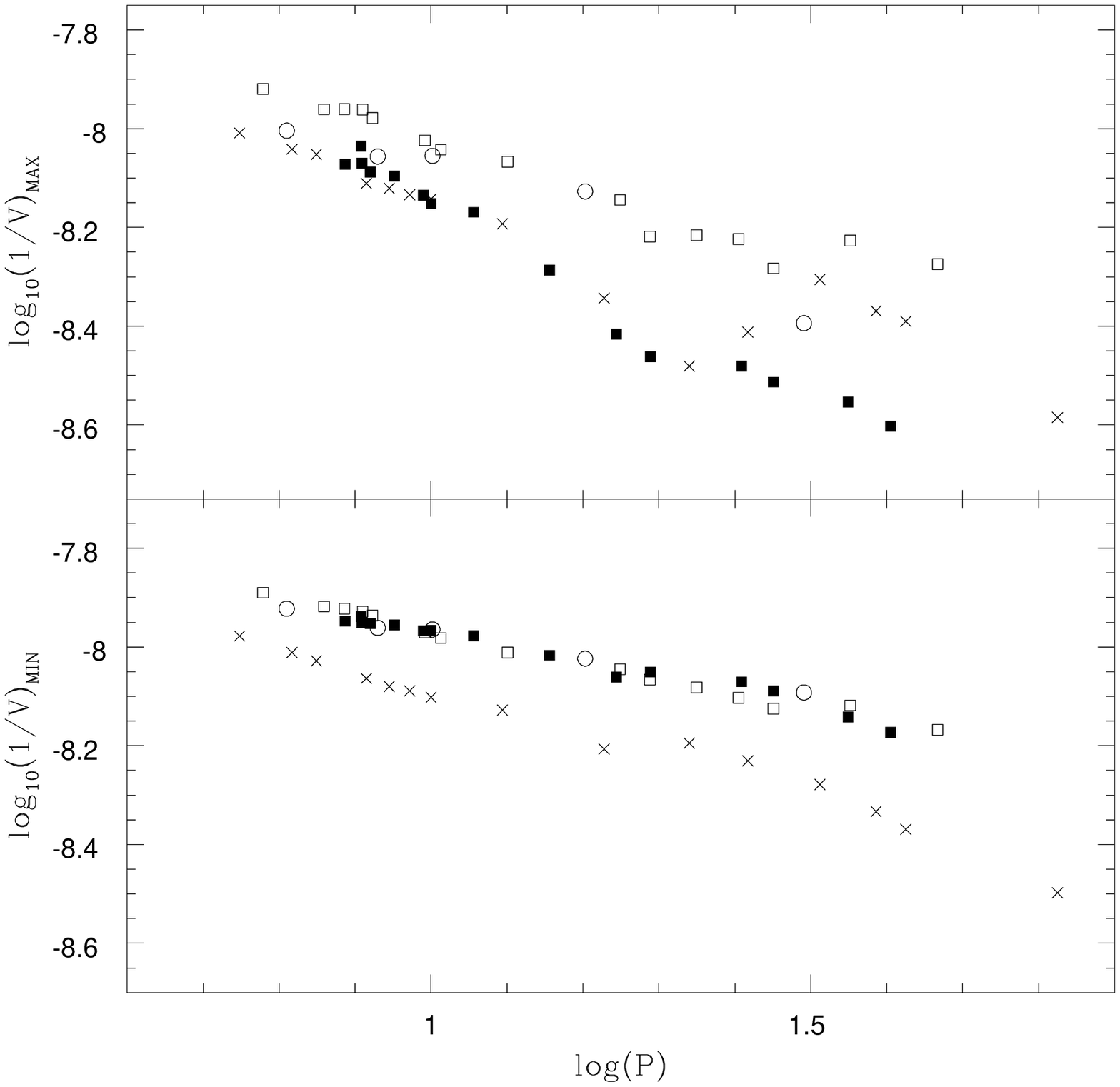}
         \epsfxsize=7.5cm \epsfbox{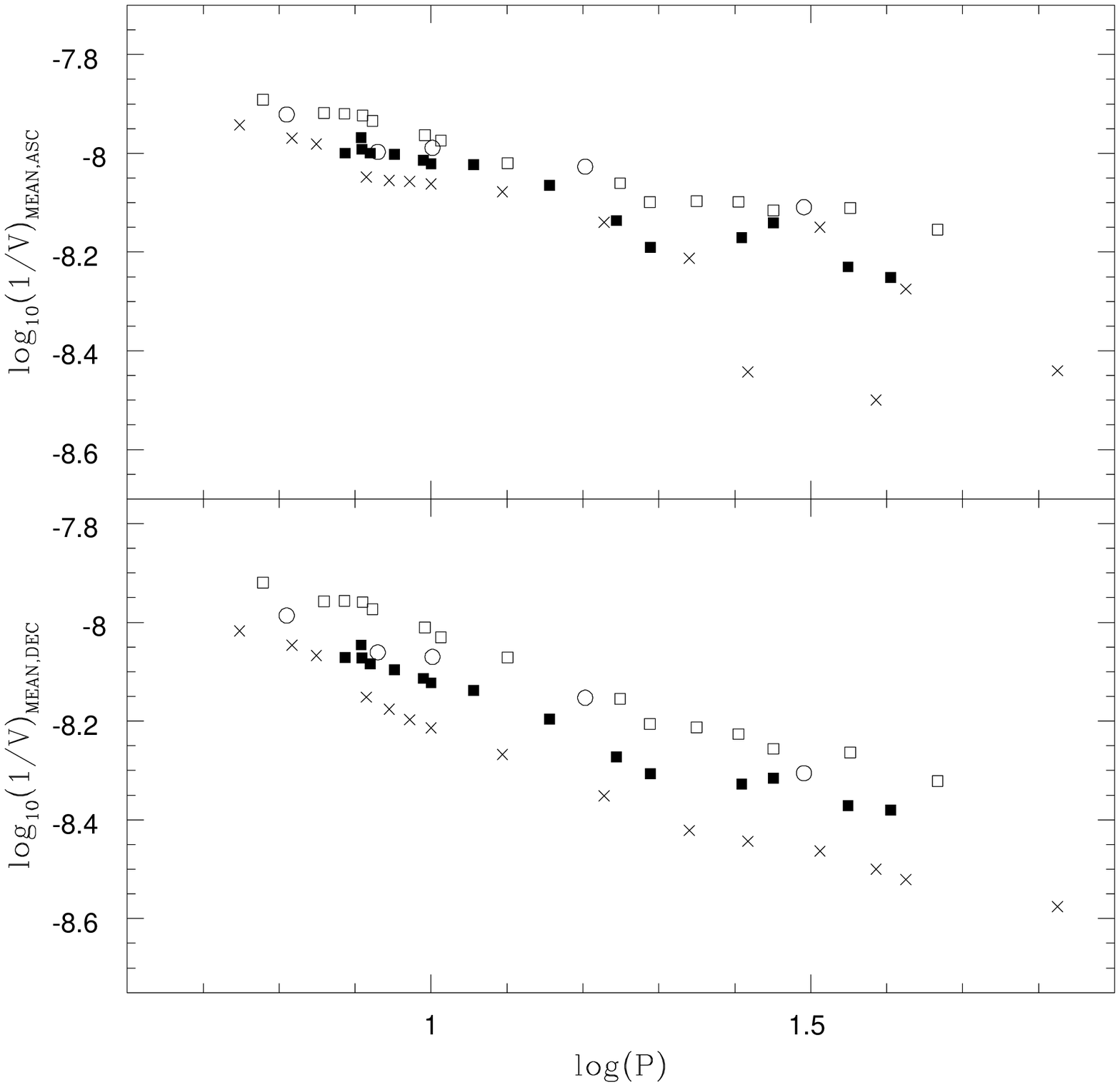}}
       \vspace{0cm}
       \caption{Log of the photospheric density plotted against period for the pulsation models. The symbols are: open squares = LMC models with Bono ML, filled squares = LMC models with Chiosi ML; open circles = LMC models with interpolated Chiosi ML; crosses = Galactic models with Chiosi ML.}
       \label{c9density}
     \end{figure*}

     Figure \ref{c9density} graphs the density (defined as $1/V$, where $V$ is the specific volume) at the photosphere as a function of the period of the model at minimum, maximum and ascending and descending mean light. Galactic models generally tend to have the lowest density and, in particular, have significantly lower densities at minimum light than the LMC models. We note that the Galactic models always have a photospheric density lower than about $10^{-8}\mathrm{g/cm}^3$ whereas the photospheric density for the LMC models only falls below this figure after a period of 10 days. At maximum light, all long period models have a low photospheric density. What we get from this figure is that it provides some evidence that there is a difference in photospheric density between the LMC and Galactic models. Moreover, this difference appears to be consistent with what is required by our theoretical scenario: short period LMC models have a higher photospheric density than their Galactic counterparts. However, for a discussion of some caveats, see Section 8.

\section{Applications}

     We now discuss two important applications of the photosphere-HIF interaction: reddening corrections and the explanation of the observed non-linear LMC PL (and PC) relations. \citet{cod47} original interest in the spectral properties of Cepheids at maximum light was to estimate reddening. SKM used this to correct a number of reddening for Galactic Cepheids. \citet{fer94} used equation (1) and the theoretical explanation provided in SKM to derive a relation linking the colour excess to the colour at maximum light, the $V$-band amplitude and the period. Such a relation is predicted from equation (1). \citet{fer94} estimates the error with this method to be comparable to other multi-colour methods.

     A more interesting application of the HIF-photosphere interaction is to explain the recent detected non-linear LMC PL relation as presented in \citet{tam02a}, Paper I, \citet{san04} and \citet{nge05}. Paper I used the $F$-test to provide strong statistical evidence that the optical Cepheid PL relation at mean light in the LMC is non-linear around a period close to 10 days. \citet{nge05} used the MACHO and 2MASS datasets together with additional long period Cepheids from the literature to further support the existence of non-linear LMC PL relation in the optical and near infra-red wave-bands. In contrast, current data indicate that the Galactic PL relation is linear at mean light \citep{tam03,nge04}. 

\begin{figure*}
       \vspace{0cm}
       \hbox{\hspace{0.2cm}\epsfxsize=7.5cm \epsfbox{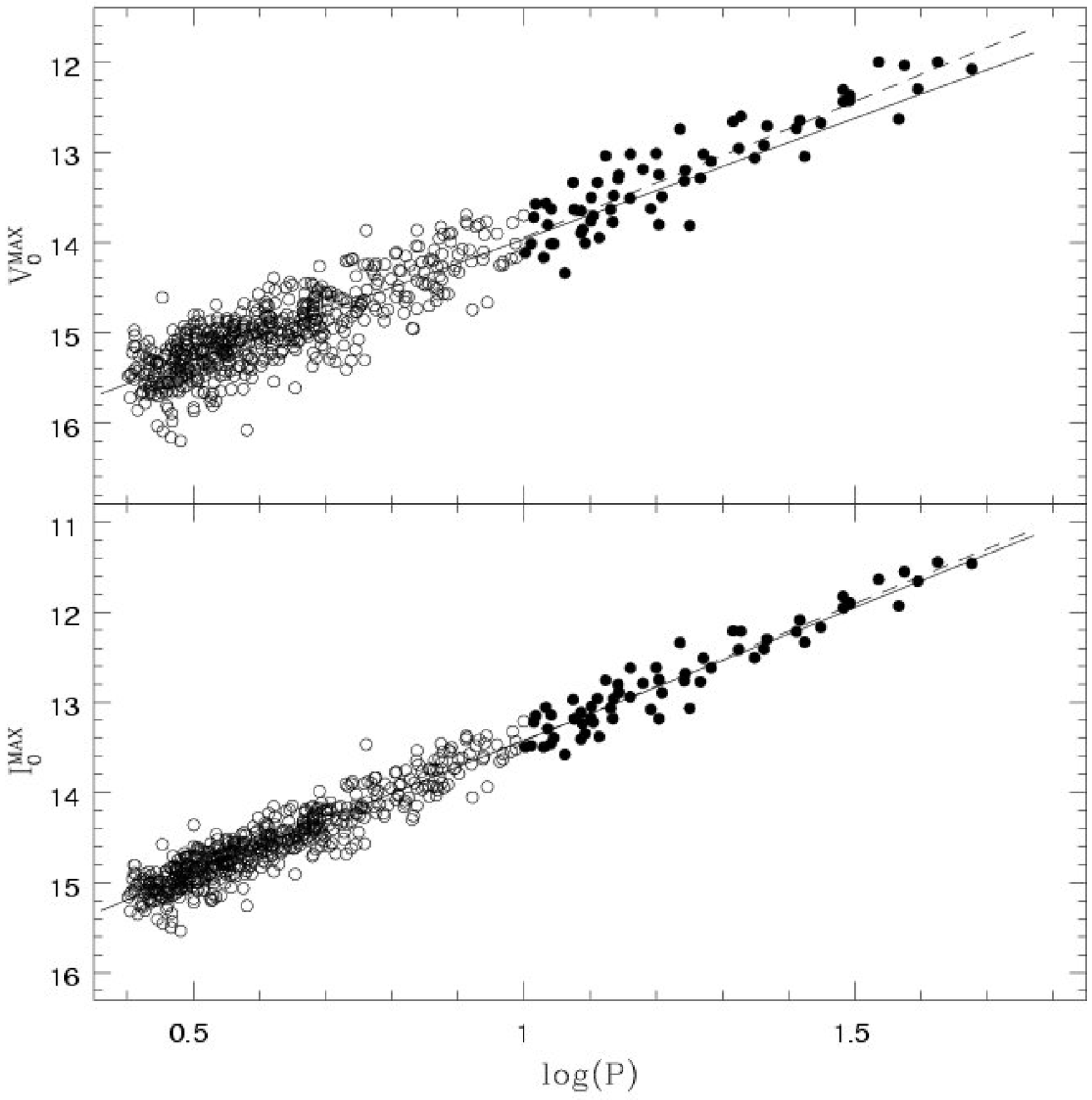}
         \epsfxsize=7.5cm \epsfbox{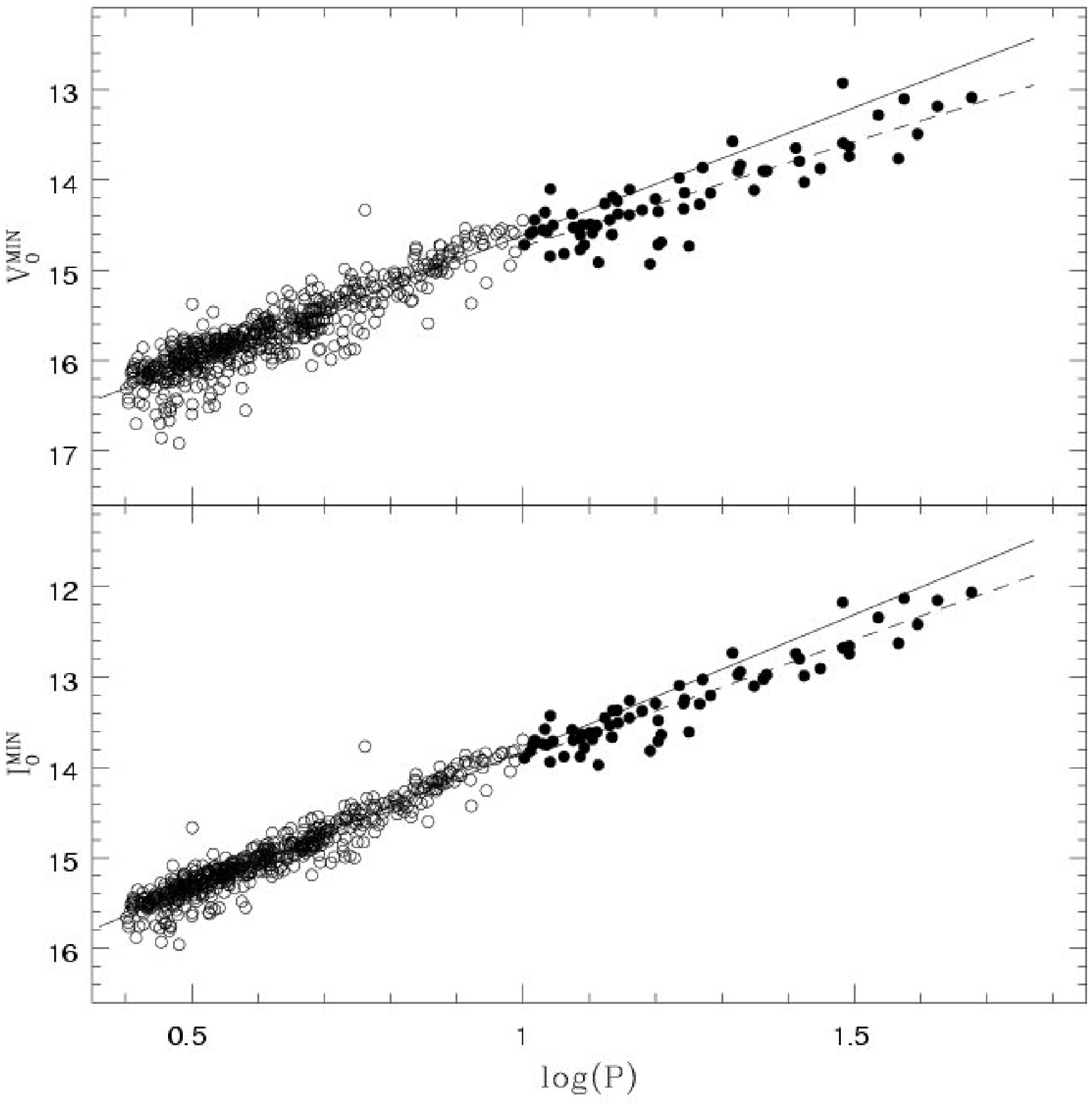}}
       \vspace{0cm}
       \caption{The LMC PL relation at maximum (left panel) and minimum (right panel) light. The upper and lower panels are for the $V$-band and the $I$-band PL relations, respectively. The open and filled circles are for short and long period Cepheids, respectively. The dashed lines are the fitted PL relations for the long period Cepheids. While the solid lines are the fitted PL relations for the short period Cepheids, and extended to the longer period range to compare with long period PL relations. }
       \label{c9plmaxmin}
     \end{figure*}

\begin{figure}
       \centering 
       \epsfxsize=7.5cm{\epsfbox{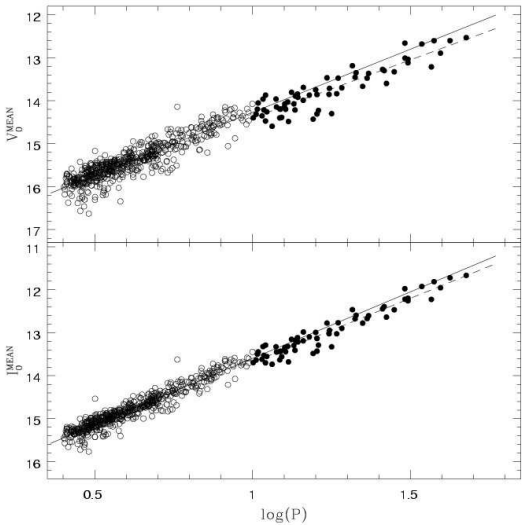}}
       \caption{Same as Figure \ref{c9plmaxmin}, but for the LMC PL relations at mean light.}
       \label{c9plmean}
     \end{figure}

     Non-linearity of the LMC PL relations can be tested using the $F$-test with the data given in Section 3. The empirical results of the fitted LMC PL relations at maximum, mean and minimum light using the updated data are presented in Table \ref{c9tabpl}. The plots of the PL relations at maximum/minimum light and at mean light are shown in Figure \ref{c9plmaxmin} \& \ref{c9plmean}, respectively. The $F$-test results for these PL relations are: $F_V(\mathrm{max,mean,min})=\{2.03,\ 8.22,\ 16.1\}$, and $F_I(\mathrm{max,mean,min})=\{0.28,\ 7.37,\ 17.9\}$. The large $F$-values for both $V$- and $I$-band PL relations at mean and minimum light strongly indicate that the PL relations at these two phases are not linear, and the data is better described with the broken (i.e, two regressions) PL relation. However, the small $F$-values at maximum light, with corresponding $p$-values of $0.13$ and $0.76$ for the $V$- and $I$-band PL(max) relations respectively, show that the null hypothesis of the $F$-test cannot be rejected (a value of $p<0.05$ and/or $F>3$ is required for doing this). Hence there is no observed break seen in the PL(max) relation and the data is consistent with single line regression. Note that the same slopes of the PL(max) relations for long period Cepheids in both bands are consistent of the finding that the PC(max) relation is flat for these Cepheids.

     \begin{table*}
       \centering
       \caption{The period-luminosity relation in the form of $m_{(V,I)}=a_{(V,I)}\log(P)+b_{(V,I)}$, and $\sigma_{(V,I)}$ is the dispersion of the relation.}
       \label{c9tabpl}
       \begin{tabular}{lcccccc} \hline
         Phase & $a_V$ & $b_V$ & $\sigma_V$ & $a_I$ & $b_I$ & $\sigma_I$\\
         \hline 
         \multicolumn{7}{c}{All, $N=641$} \\   
         Maximum & $-2.792\pm0.044$ & $16.706\pm0.031$ & 0.260 & $-2.966\pm0.029$ & $16.380\pm0.020$ & 0.170 \\
         Mean    & $-2.735\pm0.035$ & $17.088\pm0.025$ & 0.208 & $-2.963\pm0.024$ & $16.609\pm0.017$ & 0.141 \\
         Minimum & $-2.546\pm0.034$ & $17.270\pm0.024$ & 0.204 & $-2.837\pm0.023$ & $16.738\pm0.017$ & 0.140 \\
         \multicolumn{7}{c}{Long, $N=63$} \\   
         Maximum & $-3.014\pm0.180$ & $16.959\pm0.225$ & 0.257 & $-3.040\pm0.122$ & $16.469\pm0.152$ & 0.174 \\
         Mean    & $-2.698\pm0.160$ & $17.099\pm0.199$ & 0.228 & $-2.982\pm0.111$ & $16.671\pm0.139$ & 0.158 \\
         Minimum & $-2.315\pm0.171$ & $17.052\pm0.213$ & 0.243 & $-2.617\pm0.122$ & $16.512\pm0.153$ & 0.174 \\
         \multicolumn{7}{c}{Short, $N=578$} \\   
         Maximum & $-2.680\pm0.077$ & $16.640\pm0.048$ & 0.260 & $-2.946\pm0.050$ & $16.368\pm0.031$ & 0.169 \\
         Mean    & $-2.937\pm0.060$ & $17.205\pm0.038$ & 0.203 & $-3.090\pm0.041$ & $16.683\pm0.026$ & 0.138 \\
         Minimum & $-2.818\pm0.057$ & $17.429\pm0.036$ & 0.194 & $-3.030\pm0.039$ & $16.852\pm0.024$ & 0.131 \\
         \hline
       \end{tabular}
     \end{table*}

%**********************************************************
%      FIGURE: PL max/min
%**********************************************************
 
     Our tentative theoretical explanation for the non-linear nature of the LMC PL relations across a period of 10 days replies on the HIF-photosphere interaction. \citet{san58} and \citet{mad91} have established the connection between the PC and PL relations: both these relations arise from the more general PLC relation. These relations refer to quantities evaluated at mean light. The existence of such a connection relies on the period-mean density theorem, the instability strip and the Stefan-Boltzmann law. If we assume the Stefan-Boltzmann law can be applied at every phase, then it is straightforward to show that a PLC relation (though possibly with different coefficients) exists at every phase point. Thus the standard PLC relation and indeed the PC and PL relation expresses at mean light are just the averages of the same relations at different phases points. Consequently one way to understand the behavior of PLC/PL/PC relations at mean light is to understand their behavior at different phase points. What we try to do in this paper is point out some evidence from our models that shows how the changing behavior of the PC relations at different phases can, in principle, arise from a consideration of the photosphere-HIF interaction at these phases. Since the mean light PC and PL relation are the average of those at all phases, these properties can affect the PC and, as a consequence, the PL relation (via the PLC relation). In fact, the new data with superb phase resolution from such micro-lensing projects such as OGLE and MACHO demands a multiphase analysis. This approach can potentially lead to a deeper understanding of the pulsation and evolution of Cepheid variables. For example, \citet{nge05a} looked at PC relations in the Galaxy and LMC as a function of phase. They found that short and long period LMC Cepheids have a shallower and steeper slope at most pulsation phases than Galactic Cepheids respectively. 

%************************************************
%  FIGURE: PL mean
%************************************************

\section{Conclusion and Discussion}

In this paper, we have confronted updated PC and AC relations at maximum, mean and minimum light for LMC Cepheids observed by the OGLE team, and additional Cepheids from the literature, with theoretical, full amplitude pulsation models of LMC Cepheids. The observed PC and AC relations provide compelling evidence of a non-linearity or break at a period of 10 days. We also constructed theoretical Cepheid pulsation models appropriate for the LMC using the Florida pulsation codes \citep{yec98} to study the HIF-photosphere interaction. 

The empirical results presented in this paper, as well as in other papers such as \citet{san04} and \citet{nge05}, provide strong empirical evidence that the PC and PL relations for the LMC Cepheids are non-linear, in the sense described in previous sections. Issues such as extinction and a lack of long period Cepheids that may cause the non-linear LMC PL and PC relations have been addressed and argued against in Paper I, \citet{san04} and \citet{nge05}, and will not be repeated here. Other arguments against the non-linear LMC PL relation include the results presented in \citet{per04}, as the authors found no evidence for a non-linear PL relation in the LMC at $JHK$-bands. However, \citet{nge05} treated the data of \citet{per04} extensively and found, in a statistically rigorous way, that the reason why \citet{per04} found linear $JHK$ PL relations, is due to the small number of short period Cepheids ($\sim18$) in their sample. \citet{nge05} also reduce the number of OGLE/MACHO LMC Cepheids and show how the $F$-test can produce a non-significant result when the number of short/long period Cepheids become small. Instead, using the 2MASS data that are cross-correlated with MACHO Cepheids, \citet{nge05} have found that the LMC $JH$-band PL relations are non-linear\footnote{Non-linear PL relations are relatively easier to see in $V$-band than in $J$-band, as shown in \citet{nge05}.} and the $K$-band PL relation starts to become linear. \citet{nge05} also discussed why this is the case. Another argument against the non-linear PL relation is that the PL relation should be universal, as found in \citet{gie05}. We argue that their results are based on a handful of Cepheids ($\sim15$) and on short periods Cepheids in a cluster whose membership to the LMC is in question. Their shallower Galactic PL relation based on the revised infra-red surface brightness method also contradicts the steeper Galactic PL relation based on independent methods from open cluster main-sequence fitting \citep{tam03,nge04,san04}.

It is worthwhile to point out that our sample selection does not affect the detection on non-linear LMC PL relation at mean light. Since the mean magnitudes of a Cepheid light curve is less affected by our constrains on selecting the Cepheids with good light curves, we can use the published (reddening corrected) mean magnitudes to test the non-linear LMC PL relation. The anonymous referee kindly provided a large sample of LMC Cepheids that combined the published mean $V$-band magnitudes from the OGLE, \citet{seb02} and \citet{cal91} datasets. There are a total of 115 long period Cepheids in this sample and the $F$-test still return a significant detection of the non-linear LMC PL relation. The OGLE+\citet{seb02} combined data also give very similar results. Similar tests have also been done in \citet{nge05} by using the MACHO data alone and the MACHO+\citet{seb02} combined data. The non-linear LMC PL relation is still present from the $F$-test results on these two datasets. Therefore we believe our sample selection does not affect the detection of the non-linear LMC PL relation. The detection of non-linear LMC PL relation from totally independent OGLE and MACHO data, using totally independent reddening estimates, suggested that this non-linearity is real and our paper is the first attempt to theoretically explain this non-linearity in terms of the HIF-photosphere interaction.

Due to small number of LMC models, it is impossible to derive the theoretical PC and AC relations with a small error on the slope and compare directly to the empirical relations. However, these LMC models can be qualitatively compared to the observations by converting some physical quantities to the observable quantities and vice versa, such as the temperature-colour conversion. Hence we compared our model light curves to the observations in terms of theoretical PC and AC relations at the phases of maximum, mean and minimum light and also in terms of the Fourier parameters from theoretical light curves with observations. The theoretical quantities from the models generally agree with the observations, but it was found out that these models tend to have smaller amplitudes and (hence) the temperature is cooler at maximum light than the real Cepheids. Though our models have some drawbacks in this comparison, our main interest is in comparing the interaction of the photosphere and HIF as a function of phase with similar results presented in Paper II for Galactic Cepheid pulsation models. The aim is {\it not} to compare our models rigorously with observations but rather to study models which match observations reasonably well in the context of the theoretical framework described in previous sections and in Paper I \& II. Nevertheless we argued that the qualitative nature of the photosphere-HIF interaction is not seriously affected by these problems. 

Our postulate is that at certain phases, this interaction can affect the PC relation due to the properties of the Saha ionization equation: specifically for reasonably low densities in Cepheid envelopes, hydrogen ionizes at a temperature that is almost independent of period. Consequently, when the photosphere is located at the base of the HIF, the photospheric temperature and hence the colour is almost independent of period. However, when this engagement occurs, but the density is greater, then the temperature at which hydrogen ionizes again becomes sensitive to global surroundings and hence on period. When the photosphere is not engaged with the HIF in this way, its temperature is again dependent on period and global stellar parameters.

For Galactic Cepheids, this HIF-photosphere interaction occurs mainly at maximum light for Cepheids with $\log P > 0.8$ (Paper II). At minimum light, there is a strong correlation between the HIF-photosphere distance and period leading to a definite AC relation at minimum light for Galactic Cepheids (SKM, Paper I \& II). In this paper, we have found tentative evidence that, for short period LMC models which match observations in the period-color plane, the HIF-photosphere interaction occurs at most phases but at densities which are too high to produce a flat PC relation. Why would these short period LMC Cepheids be different in this regard to short period Galactic Cepheids? One possibility could be that this is partly because these LMC Cepheids are hotter than their Galactic counterparts \citep{kan04,san04}. The HIF-photosphere are disengaged for most of the pulsation cycle for long period LMC Cepheids. This happens because as the period increases, so does the $L/M$ ratio which pushes the HIF further inside the mass distribution. When the HIF-photosphere are disengaged in this way, the photospheric temperature is more dependent on density and hence on period. The change is sudden because the HIF-photosphere are either engaged or they are not. This can lead to a sudden change in the PC relation at 10 days as shown by the observations \citep{tam02a,kan04,san04,nge05}. However, at maximum light the HIF-photosphere are engaged at low densities for long period LMC Cepheids leading to the observed flat PC relation for these stars. Taken together with equation (1), this theoretical scenario is consistent with the observed PC-AC behavior described in Paper I and in this study. The anonymous referee has noted that these suggestions about photospheric density can be tested by spectroscopic means.

We now enumerate some caveats to our argument that could be addressed in future papers.

\begin{enumerate}
\item Since the SMC PC relation at mean light is linear (e.g., Paper I), how do SMC (i.e., metal-poor) models fit into the theoretical scenario outlined in this paper and Paper II, if at all? This is a difficult question and its full answer is beyond the scope of this paper. However, as the metallicity decreases, we do note that the SMC has a different ML relation to the LMC and Galaxy and so does the temperatures associated with the instability strip. These will change the relative location of the HIF and photosphere \citep{kan95,kan96} and possibly alter the phase at which they interact. Further the amplitudes for SMC Cepheids are smaller due to the lower metallicity \citep{pac00}. This will also affect the HIF-photosphere interaction. One difference which can be consistent with this is the fact that the PC relation at maximum light in the SMC is not flat (see Paper I) but it is the case for the Galaxy and LMC PC relations. This indicates that at maximum light, there is less interaction between the HIF and photosphere at low densities. This leads to an observed linear PC relation at mean light for the SMC Cepheids. These will be investigated further in a future paper in this series.

\item Could the well-known Hertzsprung progression play any part in causing the observed changes in the Galactic and LMC PC relations? 

\item It may also be that higher order overtones becoming unstable or stable, though with the fundamental mode still being dominant, may also  have an impact on the PC relation in some as yet unknown way (Paper II).

\item The behavior of short period LMC Cepheids still needs to be understood, for example, what causes the difference between the bottom left panels of Figures \ref{c9deltalmc} and \ref{c9delta}? That is, why is it that for short period Galactic/LMC Cepheids, the HIF-photosphere are disengaged/engaged? Our experience suggests that constructing short period full amplitude fundamental mode Cepheids requires more care than the long period case because the first overtone has a non-negligible growth rate. Because of this we feel a thorough study of these short period Cepheids merits a separate paper.

\item Would more advanced pulsation codes which, for example, can match the observed amplitudes and which contain a more accurate model of time dependent turbulent convection, yield similar results, especially for Figure \ref{c9delta}? Could such codes fare better in modeling short period LMC Cepheids?

\end{enumerate} 
          
\section*{acknowledgments}
     
SMK acknowledges support from  HST-AR-10673.04-A. We thank an anonymous referee for several useful suggestions and providing the data for our testing. We would also like to thank E. Antonello, R. Buchler \& J. Kwan for useful discussions, and R. Bell \& M. Marengo for the discussion regarding the atmosphere fits. 
 
%************************************
%  REFERENCE
%************************************

\end{document}